\documentclass[aps, prd, reprint,
superscriptaddress, amsmath, amssymb, floatfix,
showpacs, subfigure]{revtex4-2}

\usepackage[utf8]{inputenc}
\usepackage{graphicx}
\usepackage{dcolumn}
\usepackage{bm}
\usepackage{lineno}
\usepackage{float}
\usepackage{xspace}
\usepackage[inline]{enumitem}
\usepackage{overpic}
\usepackage{xcolor}
\usepackage{placeins}
\usepackage{subfigure}
\usepackage{booktabs}
\usepackage{longtable}
\usepackage[nowatermark]{fixmetodonotes}
\usepackage{tabulary}
\usepackage{isotope}
\usepackage{array}
\newcolumntype{L}[1]{>{\raggedright\let\newline\\\arraybackslash\hspace{0pt}}p{#1}}
\newcolumntype{C}[1]{>{\centering\let\newline\\\arraybackslash\hspace{0pt}}p{#1}}
\newcolumntype{R}[1]{>{\raggedleft\let\newline\\\arraybackslash\hspace{0pt}}p{#1}}

\usepackage{lineno}

\usepackage[colorlinks, urlcolor=blue, citecolor=., menucolor=.,
  anchorcolor=., linkcolor=., runcolor=.]{hyperref}

\begin{document}
\title{New CC0$\pi$ GENIE Model Tune for MicroBooNE} 


\newcommand{\Bern}{Universit{\"a}t Bern, Bern CH-3012, Switzerland}
\newcommand{\BNL}{Brookhaven National Laboratory (BNL), Upton, NY, 11973, USA}
\newcommand{\UCSB}{University of California, Santa Barbara, CA, 93106, USA}
\newcommand{\Cambridge}{University of Cambridge, Cambridge CB3 0HE, United Kingdom}
\newcommand{\CIEMAT}{Centro de Investigaciones Energ\'{e}ticas, Medioambientales y Tecnol\'{o}gicas (CIEMAT), Madrid E-28040, Spain}
\newcommand{\Chicago}{University of Chicago, Chicago, IL, 60637, USA}
\newcommand{\Cincinnati}{University of Cincinnati, Cincinnati, OH, 45221, USA}
\newcommand{\CSU}{Colorado State University, Fort Collins, CO, 80523, USA}
\newcommand{\Columbia}{Columbia University, New York, NY, 10027, USA}
\newcommand{\Edinburgh}{University of Edinburgh, Edinburgh EH9 3FD, United Kingdom}
\newcommand{\FNAL}{Fermi National Accelerator Laboratory (FNAL), Batavia, IL 60510, USA}
\newcommand{\Granada}{Universidad de Granada, Granada E-18071, Spain}
\newcommand{\Harvard}{Harvard University, Cambridge, MA 02138, USA}
\newcommand{\IIT}{Illinois Institute of Technology (IIT), Chicago, IL 60616, USA}
\newcommand{\KSU}{Kansas State University (KSU), Manhattan, KS, 66506, USA}
\newcommand{\Lancaster}{Lancaster University, Lancaster LA1 4YW, United Kingdom}
\newcommand{\LANL}{Los Alamos National Laboratory (LANL), Los Alamos, NM, 87545, USA}
\newcommand{\Manchester}{The University of Manchester, Manchester M13 9PL, United Kingdom}
\newcommand{\MIT}{Massachusetts Institute of Technology (MIT), Cambridge, MA, 02139, USA}
\newcommand{\Michigan}{University of Michigan, Ann Arbor, MI, 48109, USA}
\newcommand{\Minnesota}{University of Minnesota, Minneapolis, MN, 55455, USA}
\newcommand{\NMSU}{New Mexico State University (NMSU), Las Cruces, NM, 88003, USA}
\newcommand{\Oxford}{University of Oxford, Oxford OX1 3RH, United Kingdom}
\newcommand{\Pitt}{University of Pittsburgh, Pittsburgh, PA, 15260, USA}
\newcommand{\Rutgers}{Rutgers University, Piscataway, NJ, 08854, USA}
\newcommand{\SLAC}{SLAC National Accelerator Laboratory, Menlo Park, CA, 94025, USA}
\newcommand{\SDSMT}{South Dakota School of Mines and Technology (SDSMT), Rapid City, SD, 57701, USA}
\newcommand{\Maine}{University of Southern Maine, Portland, ME, 04104, USA}
\newcommand{\Syracuse}{Syracuse University, Syracuse, NY, 13244, USA}
\newcommand{\TelAviv}{Tel Aviv University, Tel Aviv, Israel, 69978}
\newcommand{\Tennessee}{University of Tennessee, Knoxville, TN, 37996, USA}
\newcommand{\UTA}{University of Texas, Arlington, TX, 76019, USA}
\newcommand{\Tufts}{Tufts University, Medford, MA, 02155, USA}
\newcommand{\VTech}{Center for Neutrino Physics, Virginia Tech, Blacksburg, VA, 24061, USA}
\newcommand{\Warwick}{University of Warwick, Coventry CV4 7AL, United Kingdom}
\newcommand{\Yale}{Wright Laboratory, Department of Physics, Yale University, New Haven, CT, 06520, USA}

\affiliation{\Bern}
\affiliation{\BNL}
\affiliation{\UCSB}
\affiliation{\Cambridge}
\affiliation{\CIEMAT}
\affiliation{\Chicago}
\affiliation{\Cincinnati}
\affiliation{\CSU}
\affiliation{\Columbia}
\affiliation{\Edinburgh}
\affiliation{\FNAL}
\affiliation{\Granada}
\affiliation{\Harvard}
\affiliation{\IIT}
\affiliation{\KSU}
\affiliation{\Lancaster}
\affiliation{\LANL}
\affiliation{\Manchester}
\affiliation{\MIT}
\affiliation{\Michigan}
\affiliation{\Minnesota}
\affiliation{\NMSU}
\affiliation{\Oxford}
\affiliation{\Pitt}
\affiliation{\Rutgers}
\affiliation{\SLAC}
\affiliation{\SDSMT}
\affiliation{\Maine}
\affiliation{\Syracuse}
\affiliation{\TelAviv}
\affiliation{\Tennessee}
\affiliation{\UTA}
\affiliation{\Tufts}
\affiliation{\VTech}
\affiliation{\Warwick}
\affiliation{\Yale}

\author{P.~Abratenko} \affiliation{\Tufts} 
\author{R.~An} \affiliation{\IIT}
\author{J.~Anthony} \affiliation{\Cambridge}
\author{L.~Arellano} \affiliation{\Manchester}
\author{J.~Asaadi} \affiliation{\UTA}
\author{A.~Ashkenazi}\affiliation{\TelAviv}
\author{S.~Balasubramanian}\affiliation{\FNAL}
\author{B.~Baller} \affiliation{\FNAL}
\author{C.~Barnes} \affiliation{\Michigan}
\author{G.~Barr} \affiliation{\Oxford}
% ...

\collaboration{The MicroBooNE Collaboration}
\thanks{microboone\_info@fnal.gov}\noaffiliation

\date{\today}

\begin{abstract}
Obtaining a high-quality interaction model with associated uncertainties is essential for neutrino experiments studying oscillations, nuclear scattering processes, or both.  As a primary input to the MicroBooNE experiment's next generation of neutrino cross-section measurements and its flagship investigation of the MiniBooNE low-energy excess, we present a new tune of the charged-current pionless (CC0$\pi$) interaction cross section via the two major contributing processes -- charged-current quasielastic and multi-nucleon interaction models -- within version 3.0.6 of the GENIE neutrino event generator. Parameters in these models are tuned to muon neutrino CC0$\pi$ cross-section data obtained by the T2K experiment, which provides an independent set of neutrino interactions with a neutrino flux in a similar energy range to MicroBooNE's neutrino beam.  Although the fit is to muon neutrino data, the information carries over to electron neutrino simulation because the same underlying models are used in GENIE.
A number of novel fit parameters were developed for this work, and the optimal parameters were chosen from existing and new sets. We choose to fit four parameters that have not previously been constrained by theory or data.  Thus, this will be called a theory-driven tune.  The result is an improved match to the T2K CC0$\pi$ data with more well-motivated uncertainties based on the fit. 
\end{abstract}

\maketitle

\section{Introduction}

A fundamental challenge in a large variety of accelerator experiments is to
have an accurate Monte Carlo simulation of the apparatus.  For neutrino experiments, a
key aspect is the neutrino interaction modeling~\cite{nustec-review}. For
MicroBooNE~\cite{acciarri_design_2017}, this is true both for the low-energy
excess (LEE) search~\cite{PRLeLEE,PeLEE,DLeLEE,WCeLEE,gLEE} (based on the findings of
MiniBoone~\cite{MiniBooNE:2012meu}) and neutrino-argon scattering cross-section
measurements. MicroBooNE uses a liquid argon target exposed to Fermilab's
Booster Neutrino Beam (BNB), which has a mean neutrino energy of
0.8~GeV~\cite{PhysRevD.79.072002}. This makes the LEE search sensitive to
nuclear effects such as multi-nucleon correlations and medium corrections.
Cross-section measurements require a model that can provide a reliable estimate
of the contribution of background events in a selection.  Although selection
cuts that decrease the number of background events and data-driven estimations
are highly desirable, problems with background very often remain and the
model-data correspondence must be close enough to trust the efficiency
estimation. MicroBooNE's LEE search, on the other hand, has the critical
requirement of a model that provides a baseline estimate for most non-LEE
contributions to the event yields and estimated uncertainties.  The model also
correlates uncertainties between different samples of selected events.
Achieving significant improvement in the understanding of both the central-value model and its related uncertainties is
the most important goal of this work.

The MicroBooNE data set contains a significant number of charged-current quasielastic
(CCQE) interactions, as well as charged-current \emph{2 particle, 2 hole} (CC2p2h) interactions in
which the neutrino interacts with a correlated pair of nucleons.  Resonant
(RES) interactions, particularly resonant pion production, also form a
considerable portion of the interactions collected by MicroBooNE (although these
interactions tend to be at the highest energies in the neutrino flux and therefore of 
less concern for MicroBooNE's electron-like LEE searches) and deep inelastic
scattering (DIS) also occurs in smaller proportions at even higher energies.
MicroBooNE has adopted GENIE~\cite{GENIE} v3.0.6 G$18\_10a\allowbreak
\_02\allowbreak \_11a$ as its core event generator at this time.  This version
of GENIE uses new models from the Valencia
group~\cite{Nieves:2011yp,NievesCCQE,Gran:2013kda} for CCQE and CC2p2h
processes.  The introduction of new models presents the problem of choosing
features of the model to test.  We develop a theory-driven set of parameters
for this purpose where the parameters are chosen with respect to the lack of
understanding in the underlying theory.

The treatment of nuclear dependence (or $A$-dependence) is important to this work because the existing data are largely for carbon targets and the MicroBooNE detector is almost totally composed of argon.  Because GENIE is required to simulate interactions with the wide variety of nuclei used in experiments, it needs a simple nuclear model.  Neutrinos interact weakly within nuclei and linear dependence on A is then used.  In GENIE, corrections come in two ways.  All hadrons are subject to final state interactions which have a basic $\sim A^{2/3}$ dependence on nucleus.  Corrections for the binding energy of the struck nucleon~\cite{Bodek:1980ar} and medium dependence in FSI~\cite{Pandharipande:1992zz} are applied; both have weak A dependence.  All medium corrections have a smooth dependence on A and are determined in fits to inclusive and semi-inclusive electron scattering data.  

In this article, we present a tune of the GENIE~\cite{GENIE} v3.0.6
G18$\_10a\allowbreak \_02\allowbreak \_11a$ CCQE and CC2p2h models to T2K
CC0$\pi$ cross-section data~\cite{t2k2016}. The main purpose of this work is to
provide an important input to the $\nu_e$-focused MicroBooNE LEE
analyses~\cite{PRLeLEE,WCeLEE,PeLEE,DLeLEE}. One of the primary goals there is
to use the $\nu_\mu$ data to constrain the $\nu_e$ model with robust
uncertainties and covariances. The goal is to propagate features properly
across channels and this fit is the basis for that exercise.  

Although we tune underlying model parameters of these specific processes, the
aim of this work is to improve the prediction of the overall CC0$\pi$
cross section, and thereby improve the model predictions and uncertainties used
in MicroBooNE analyses.  The signal choice of CC0$\pi$ means no mesons in the
final state; experiments use this choice because it has less model dependence
than true CCQE (which cannot be directly measured in a neutrino experiment). It includes CCQE, CC2p2h, and some pion production (where the pion is
absorbed in the residual nucleus) events.  

The choice of the data set to study
here is an important decision that is particular to the MicroBooNE experiment. Tuning to external data (i.e. data from another experiment) allows us to avoid any potential biases from double-fitting to MicroBooNE data in any subsequent analyses. 
Although choices of target and beam energy must be considered, the emphasis is
on external data sets with similar beam energy because the number and complexity of allowed interaction channels grow as the beam energy increases.  The T2K~\cite{t2k2016,PhysRevD.97.012001,t2k2018,PhysRevD.101.112001,PhysRevD.101.112004} and MiniBooNE~\cite{AguilarArevalo:2007ab,miniboone-ccqe} experiments both have CC0$\pi$ data sets with neutrino fluxes in a very similar energy range to MicroBooNE (in fact, the MiniBooNE experiment sits in the same beamline as MicroBooNE and sees an almost identical neutrino flux). MINERvA has also published
CC0$\pi$ data~\cite{PhysRevLett.124.121801,PhysRevD.101.092001,Ruterbories:2018gub,PhysRevLett.121.022504,PhysRevLett.119.082001,PhysRevD.91.071301,PhysRevLett.111.022502}, but this selection has a high proportion of RES and DIS interactions due to the larger neutrino energies. It should also be noted that all three experiments use a CH target (T2K also has some water-target data), whereas MicroBooNE uses an argon target.  ArgoNeuT data~\cite{ArgoNeuT:2016CC0pion_NUINT15,ArgoNeuT:2014back-to-back} uses an argon target, but is at a
higher energy and has significant statistical uncertainties.  

Restricting the study to neutrino energies below 2 GeV
limits consideration to the CCQE and CC2p2h models. This would imply looking at data from the T2K and MiniBooNE experiments. However, there is one more consideration: independence of the data used for tuning from other MicroBooNE simulations and analyses. 
MiniBooNE is also located in the Booster Neutrino Beamline at Fermilab, in a similar location to MicroBooNE. Both experiments see an almost-identical neutrino flux and use the same simulation to calculate the neutrino flux prediction and uncertainties. This means that the flux uncertainties included in the reported MiniBooNE cross-section data will be extremely correlated with those used in MicroBooNE analyses. This issue of correlations between the two measurements is complicated: if not handled properly, it runs the risk of double-counting uncertainties or propagating unknown biases from incorrect or misunderstood calculations. Correctly accounting for shared simulation components between published MiniBooNE cross-section data and ongoing MicroBooNE analyses is currently infeasible, so we restrict our focus in model tuning to data from T2K. Considering T2K data alone minimizes the
risk that the data used for tuning would be correlated with subsequent MicroBooNE analysis inputs.



Nevertheless, MiniBooNE CC0$\pi$ data are still important for this work. These data have provided an important basis for understanding interaction mechanisms in this energy range, and have informed the model choices in GENIE. While we do not tune to data from either MiniBooNE or MicroBooNE,
comparisons of the nominal and tuned
GENIE predictions with these data are a powerful tool to evaluate the tuned models, and
therefore are shown in Secs.~\ref{sec:geniemodels},~\ref{sec:fitresults:minib} (MiniBooNE), and~\ref{sec:fitresults:uboone} (MicroBooNE).

This work was required because the default GENIE v3.0.6 G$18\_10a\allowbreak
\_02\allowbreak \_11a$ model configuration was found to under-predict
cross-section data from various experiments for charged-current interactions
that produce no charged or neutral pions above detection threshold (the CC0$\pi$ channel, which is dominated by CCQE and CC2p2h
interactions). This is discussed further in Sec.~\ref{sec:geniemodels}.
The features of GENIE that are important for this fitting exercise are also
presented in Sec.~\ref{sec:geniemodels}.  

The data chosen for tuning and
selection of parameters needed to properly tune the model for MicroBooNE are
described in Sec.~\ref{sec:fitmeth}. Fit results are presented in
Sec.~\ref{sec:fitresults} along with comparisons to other relevant data. A full description of MicroBooNE's adopted set of neutrino cross-section model uncertainties is given in Sec.~\ref{sec:systematics}, and the conclusions from this study are summarized in Sec.~\ref{sec:conclusions}.

\section{GENIE models}
\label{sec:geniemodels}

The latest version of GENIE~(v3)~\cite{GENIEepj} has a variety of model sets
available to allow users to make the best choice for their particular analysis.
The GENIE v3.0.6 G$18\_10a\allowbreak \_02\allowbreak \_11a$ model set includes
the full Valencia model~\cite{Nieves:2011yp,NievesCCQE,Gran:2013kda} for the
local Fermi gas nucleon momentum distribution, CCQE, and CC2p2h interactions.
This adds the RPA correction (random phase approximation which is a description
of long-range nucleon-nucleon correlations) and the Coulomb interaction of the
outgoing muon. It also includes improved agreement with a significantly
expanded data set for the $A$-dependence of final state interactions (FSI)
along with updated form factors~\cite{Graczyk:2008zz} and diagrams for the pion
production process~\cite{Berger:2007rq,Kuzmin:2003ji,Nowak:2009se}. A new tune
to neutrino-proton and neutrino-deuterium cross-section
data~\cite{FreeNucleonTune} has been applied. These features represent
significant enhancements for neutrino energies sampled by MicroBooNE over the
historical default model from GENIE v$2.12.2$ (used by previous MicroBooNE
analyses~\cite{PhysRevLett.123.131801,uB_CCQE,uB_CCnp,uB_nue_NuMi}).

The Valencia models as implemented in GENIE v3 are
somewhat different than those in the original publications~\cite{Nieves:2011yp,NievesCCQE,Gran:2013kda} outlining the Valencia models. The CC2p2h
model had a restriction on the 3-momentum transfer~\cite{Gran:2013kda} which
has little effect at lower neutrino energies.  The CCQE model~\cite{NievesCCQE}
applied a simple binding energy correction by subtracting it from the muon
energy. GENIE v3 used a more detailed method of decreasing the effective mass of
the struck nucleon~\cite{Bodek:1980ar,GENIEepj}. This feature will have an
important impact on the analysis presented here.

\begin{figure}[hbt]
    \centering
     \includegraphics[width=0.45\textwidth]{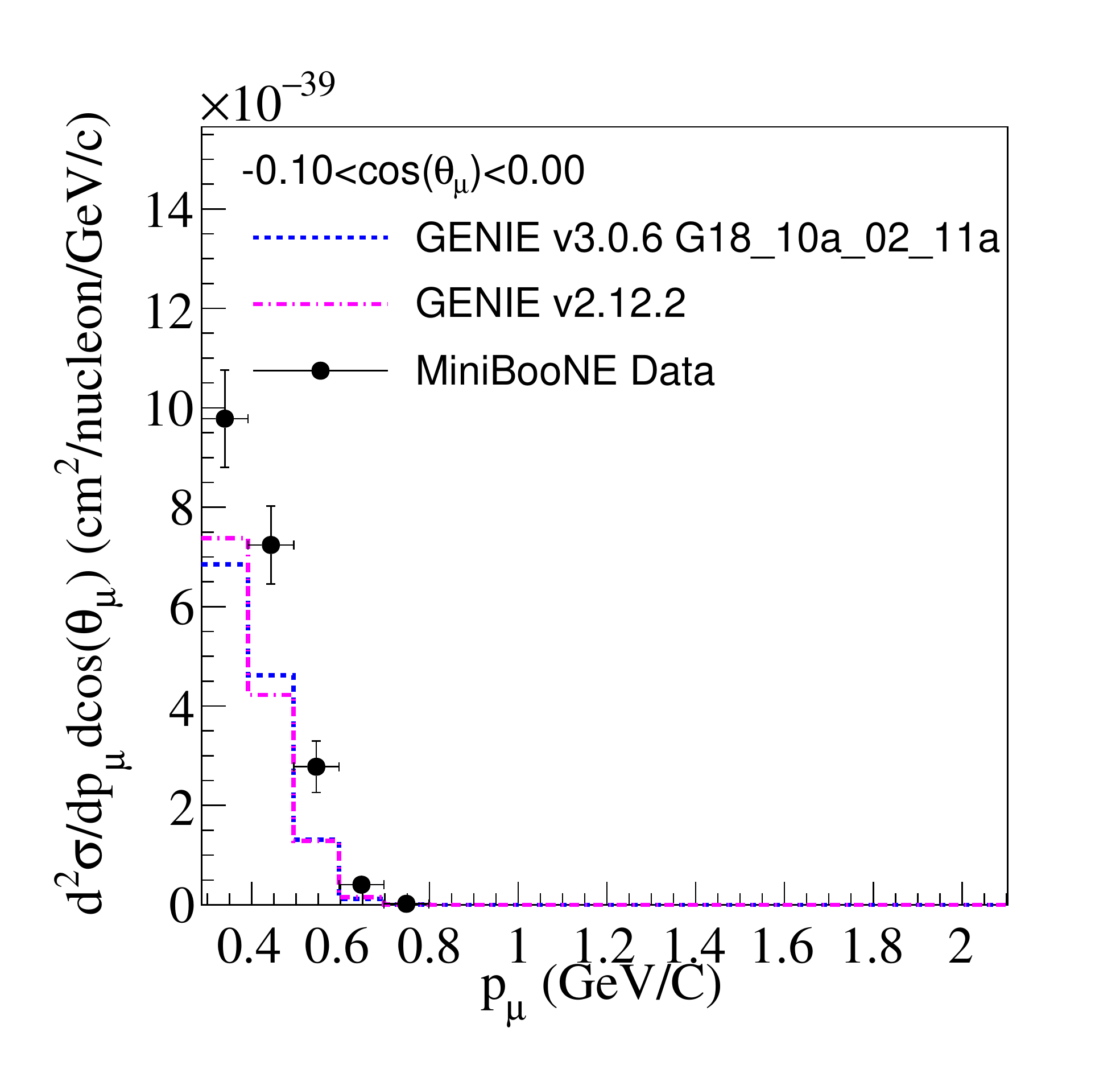}
   \includegraphics[width=0.45\textwidth]{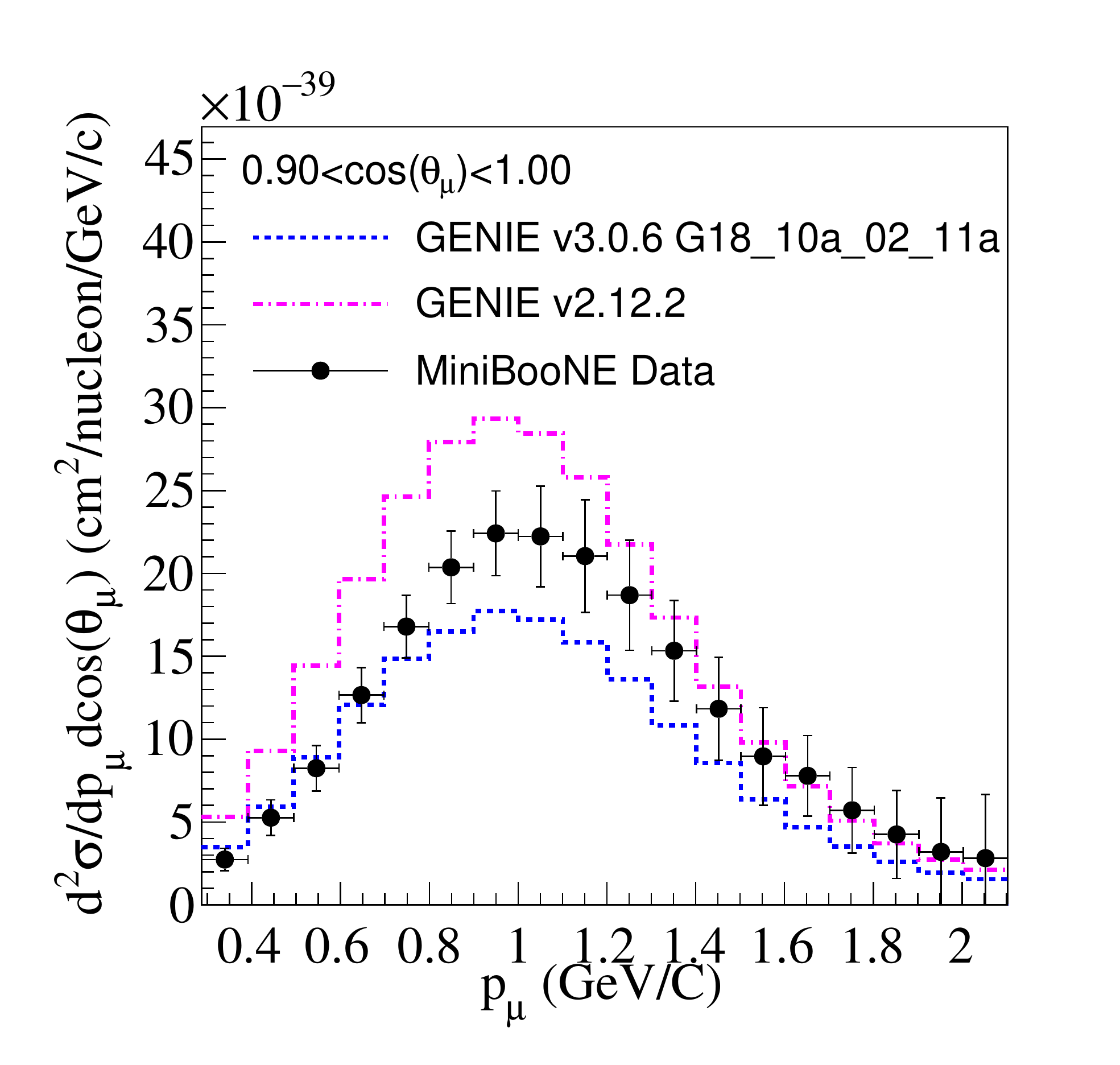}
    \caption{MiniBooNE flux-averaged CC0$\pi$-like differential cross
section~\cite{miniboone-ccqe} for muon kinetic energy $p_{\mu}$ compared with
GENIE v2.12.2 and v3.0.6 G$18\_10a\allowbreak \_02\allowbreak \_11a$
simulations for $-0.1 < \cos\theta_{\mu} < 0.0$ (top) and
$0.9 < \cos\theta_{\mu} < 1.0$ (bottom). The original data release is in terms of muon kinetic energy. Uncertainties on the data points are the shape uncertainties reported by the collaboration.  Uncertainties reported include a  10.7\% normalization error and a $\chi^2$ analysis is presented in Table~\ref{tab:mChi2_intro}.  These bins show the underprediction
of GENIE v3.0.6 G$18\_10a\allowbreak \_02\allowbreak \_11a$ compared to data
for forwards-going and backwards-going muons.}
    \label{fig:intro:mB}
\end{figure}

The updated models have been validated against both inclusive and exclusive
bubble-chamber experiments, and provide a significantly improved description of
CC$0\pi$ cross-section data from MiniBooNE~\cite{miniboone-ccqe} compared to
the GENIE v$2.12.2$ models.  The relativistic Fermi gas nuclear model and
Llewellyn-Smith QE model used in GENIE v$2.12.2$ are more appropriate for
measurements at a higher beam energy than the BNB energies.
Fig.~\ref{fig:intro:mB} shows a comparison of MiniBooNE CC0$\pi$
data~\cite{miniboone-ccqe} with simulations from these two models. GENIE v3.0.6
G$18\_10a\allowbreak \_02\allowbreak \_11a$ more closely matches the data and
will be the basis for the fit we describe in this document. However, it is also
clear that GENIE v3.0.6 G$18\_10a\allowbreak \_02\allowbreak \_11a$
underpredicts the data in both angular bins shown ($-0.1 < \cos(\theta_{\mu}) <
0.0$ and $0.9 < \cos(\theta_{\mu}) < 1.0$) where $\theta_\mu$ is the lab angle
of the muon with respect to the best estimate of the neutrino direction. Table~\ref{tab:mChi2_intro} provides a $\chi^2$ analysis of these two bins using the shape and normalization uncertainties provided by the collaboration.  A
similar underprediction can be seen when comparing to CC0$\pi$ cross-section
data from T2K~\cite{t2k2016} in Fig.~\ref{fig:t2kfits:nominal_analysis1}, as well as 
in data selections for MicroBooNE, such as that shown in
Fig.~\ref{fig:results:ubooneWC:notune}.  The underproduction of MicroBooNE data
is the primary motivation for this tuning work in the context of the LEE
search.

\begin{table*}[htb]
    \caption{$\chi^2$ for comparisons of GENIE v2.12.2 and v3.0.6 G18\_10a\_02\_11a to MiniBooNE flux-averaged CC0$\pi$-like differential cross
section data~\cite{miniboone-ccqe} for the two bins presented in Fig.~\ref{fig:intro:mB}. Following the information presented in the publication, $\chi^2$ are calculated separately for shape and normalization components, including a reported 10.7\% normalization error. Each value is then presented in the table.  No correlations between bins were included.}
    \centering
    \begin{tabular}{|c|c|c|} \hline 
    $\chi^2_{shape}/\rm{N_{bins}}$,$\chi^2_{norm}/\rm{N_{bins}}$& GENIE v3.0.6 G18\_10a\_02\_11a  & GENIE v2.12.2  \\ \hline 
   -0.10$<\rm{cos(\theta_\mu)}<$0.00    & 35.11/4, 11.41 & 34.60/4, 10.97  \\ \hline 
    0.90$<\rm{cos(\theta_\mu)}<$1.00    & 24.03/18, 3.83  &  110.46/18, 6.59  \\ \hline  
    \end{tabular}
    \label{tab:mChi2_intro}
\end{table*}


The choice to tune specifically the CCQE and CC2p2h models in this work is
motivated by several factors.  The largest difference between data and
simulation in many MicroBooNE analyses is seen at low visible energy (see, for
example, Fig.~\ref{fig:results:ubooneWC:notune}) where these interactions
dominate.  The lack of consensus about the optimal choice of these models in the
theoretical community~\cite{nustec-review,Gonzalez-Jimenez:2019ejf,%
Rodrigues:2015hik,NuWroFit} strongly drives the parameter choices in this fitting exercise. Tuning other channels (e.g. resonant pion
production) has lower priority for MicroBooNE at this time, mainly because the
contribution of these channels is less relevant for MicroBooNE's electron-like
LEE searches due to the energy range and the effective $\pi^0$ rejection
achieved by these analyses. However, the work presented here may be extended to
include tuning of other channels in the future.

\section{Fitting Method}
\label{sec:fitmeth}
Fits to T2K CC0$\pi$ cross-section data~\cite{t2k2016} are performed to tune
parameters within the GENIE v3.0.6 G18\_10a\_02\_11a CCQE and CC2p2h models
using the GENIE reweighting package v1.0.4. The main goal is to mitigate the
underprediction observed in both MicroBooNE and MiniBooNE data. The tuning was
performed using the NUISANCE software package~\cite{NUISANCE}.

\subsection{T2K Data Set}
\label{sec:t2kdata}

The choice of the single data set to include in a fit is important as the data
set must be as close as possible to the MicroBooNE data and contain a high
proportion of CCQE and CC2p2h interactions. The inclusion of more data sets
would have complicated the fit because of the incompatibilities that can be
exposed~\cite{tensions2016}. It was also important to limit the fit to data at
neutrino energies compatible with MicroBooNE because of the growing complexity
of models needed to describe data properly as the energy
increases~\cite{Stowell:2019zsh}.

We perform fits to T2K CC0$\pi$ cross-section data, published in
2016~\cite{t2k2016}. These results come from an analysis that requires an
inclusive muon measurement.  A later analysis of the same data includes results
based on proton multiplicity~\cite{t2k2018}; for these fits the 2016 inclusive
muon-based data are sufficient~\cite{t2k2016}.  The paper includes two
analyses, which both use the same data events but apply different selection
cuts and extract the cross section using different methods:
\begin{itemize}
    \item {\bf Analysis 1:} Uses a binned likelihood fit performed
simultaneously in four signal and two control regions to extract the cross
section. A 2D double-differential cross section as a function of muon momentum
(p$_{\mu}$) and angle ($\cos{\theta_{\mu}}$) is provided in the full phase
space.
    \item {\bf Analysis 2:} Uses d'Agostini unfolding to extract the cross
section. A 2D double-differential cross section as a function of muon momentum
(p$_{\mu}$) and angle ($\cos{\theta_{\mu}}$) is provided in the restricted
phase space where p$_{\mu} > 0.2$ GeV and $\cos{\theta_{\mu}} > 0.6$.
\end{itemize}

For the work presented in this article, we fit to ``Analysis 1'' from the
paper, both because of the broader phase space of the ``Analysis 1'' result and
because d'Agostini unfolding is known to produce undercoverage of
uncertainties.

\begin{figure*}
    \centering
    \includegraphics[width=\textwidth]{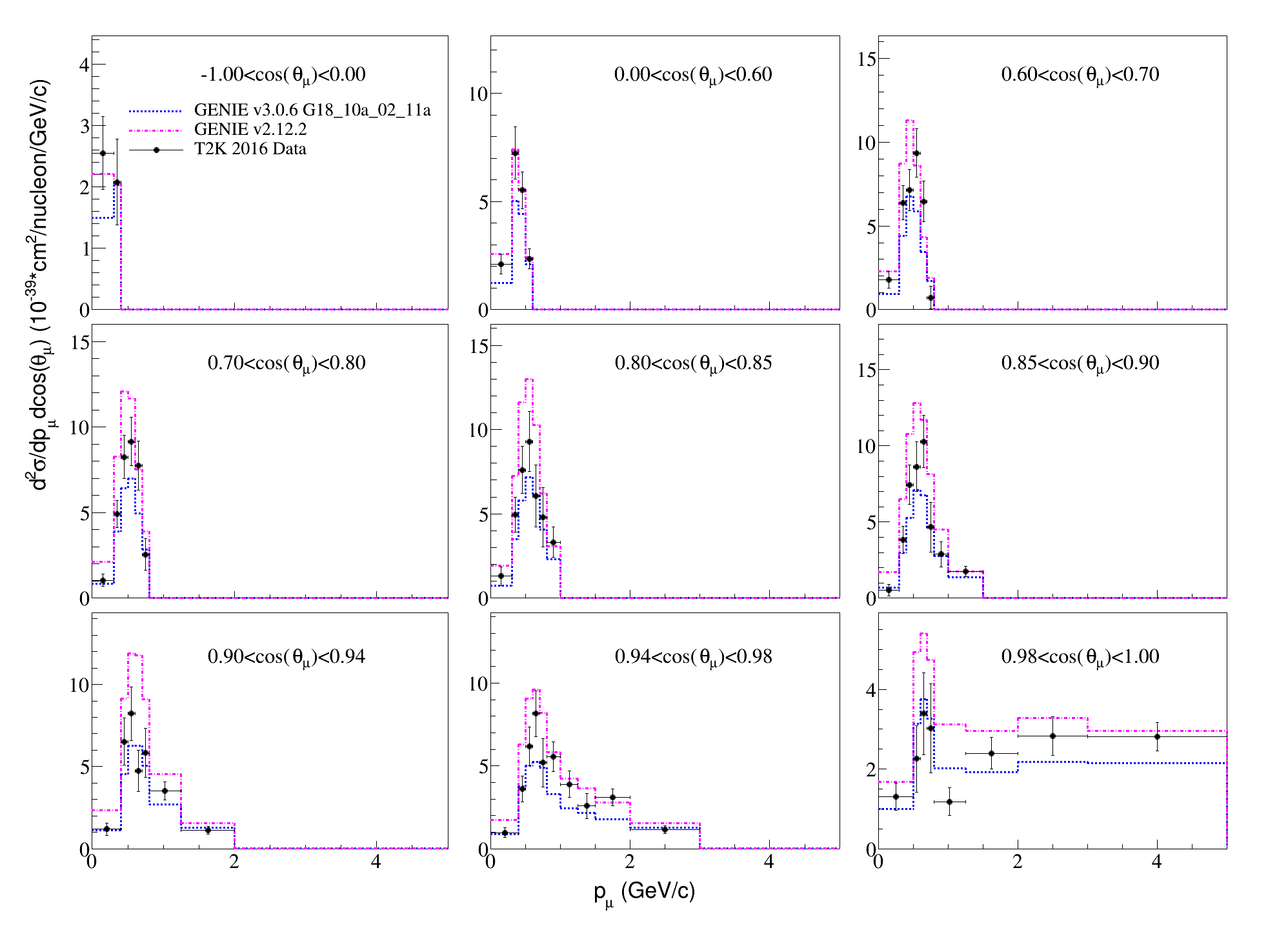}
    \caption{T2K flux-averaged CC0$\pi$ analysis 1 double-differential cross section of lepton
momentum and $\rm{cos(\theta_{\mu})}$ \cite{t2k2016} compared to nominal GENIE v3.0.6 G$18\_10a\allowbreak
\_02\allowbreak \_11a$ ($\chi_{diag}^2/\rm{N_{bins}}$=115.31/67) and GENIE v2.12.2
($\chi_{diag}^2/\rm{N_{bins}}$=284.31/67). The calculation of  $\chi_{diag}^2$ uses only the uncertainties on the diagonals of the T2K CC0$\pi$ data set. While only bins of $\rm{p_{\mu}}<$ 5~GeV are plotted, all bins with data up to $\rm{p_{\mu}}=$ 30~GeV are included in the $\chi_{diag}^2$ calculations.}
    \label{fig:t2kfits:nominal_analysis1}
\end{figure*}

Figure~\ref{fig:t2kfits:nominal_analysis1} shows the complete T2K CC0$\pi$
analysis 1 data compared to GENIE v3.0.6 G18\_10a\_02\_11a and GENIE v2.12.2.
The data are presented in a double-differential cross section as a function of
muon momentum (p$_{\mu}$) and angle ($\cos{\theta_{\mu}}$). We see the general
pattern that the more recent GENIE prediction is below the T2K data, as we have
also seen in MicroBooNE simulation.
The highest bin in muon momentum in each $\cos\theta$ bin is excluded in the
fitting procedure because very high-momentum muons are not significant in the
MicroBooNE analysis.  In addition, the very small cross sections and small
absolute uncertainties on the data in these bins can drive the fit in an
undesirable way. This reduces the number of bins in the data fit from 67
(including high-muon momentum bins) to 58.


NUISANCE allows the use of the published bin-to-bin covariance matrix in the
fitting process; this is generally regarded as the correct way to fit the data.
Unfortunately, attempts to fit the T2K data with the full covariance matrix
proved problematic, resulting in a significant and unphysical reduction of the
cross section across all of the bins. Multiple hypotheses have been considered
to explain this observation.
We note that a recent publication in which NuWro models are fit to T2K and
MINERvA data sees a similar effect when fitting the MINERvA
data~\cite{NuWroFit} and attributes it to Peelle's Pertinent Puzzle~\cite{ppp}. 

For the main result of this work, we avoided these problems with fits that use only diagonal elements of the full covariance matrix -- corresponding to the error bars on each data point drawn in
Fig.~\ref{fig:t2kfits:nominal_analysis1} -- and do not consider correlations
between bins. 
As discussed in Section~\ref{sec:fitresults}, additional tests using alternate methods~\cite{kochChi2} to include the effect of correlations were conducted to test the robustness of this approach.   

\subsection{Parameter Choice}

When using the MINUIT fitting formalism, the choices of parameters to fit are
limited because of the difficulty of finding a global minimum with reasonable
uncertainties, especially when parameters are highly correlated.  In addition,
parameters should be chosen to reflect the underlying physics.  This means
choosing parameters that clearly show a particular physical effect that is not
already constrained by data.  For example, the CC2p2h contribution is uncertain
and makes its study very useful to a variety of researchers.

The decision on parameters to avoid is also important.  Although many other fitting exercises use the Fermi momentum as a free
parameter~\cite{AguilarArevalo:2007ab}, we ignored it because it is already
well-known from electron scattering and any significant differences between
electron and neutrino probes are unlikely.  Although the MicroBooNE LEE
analysis is sensitive to protons in the final state, we do not use any FSI
parameters for protons because the T2K 2016 data~\cite{t2k2016} contains only
information on the muon. This is a limitation of this analysis; future work may
find it beneficial to add FSI parameters in a fit to measurements of proton
kinematics in the T2K data~\cite{t2k2018}.

The reweighting package provided with GENIE allows tuning of the CCQE model via
the parameter MaCCQE (which affects both shape and absolute normalization) and
tuning of the CC2p2h model via an absolute normalization parameter. These are
obvious choices to include in our tuning. We also consider a number of other
parameters that we expect might have an effect on this data set: some available
in the GENIE reweighting package and some that were developed and added to
GENIE for this work.  Sec.~\ref{sec:par:fitted} lists the parameters that were
chosen to include in the fit to T2K data, and Sec.~\ref{sec:par:notfit} details
additional new parameters that were developed for the MicroBooNE interaction
uncertainty evaluation, but ultimately not included in the fits presented here.

\subsubsection{Fit Parameters}
\label{sec:par:fitted}

\textbf{MaCCQE:} Axial mass in the dipole form factor used in the CCQE cross
section calculation. Increasing this parameter's value has the effect of
increasing the normalization of CCQE interactions, as well as changing the
shape of the cross section as a function of Q$^2$, where Q is the four-momentum
transfer to the nucleus.

\begin{figure}
    \centering
    \includegraphics[width=0.45\textwidth]{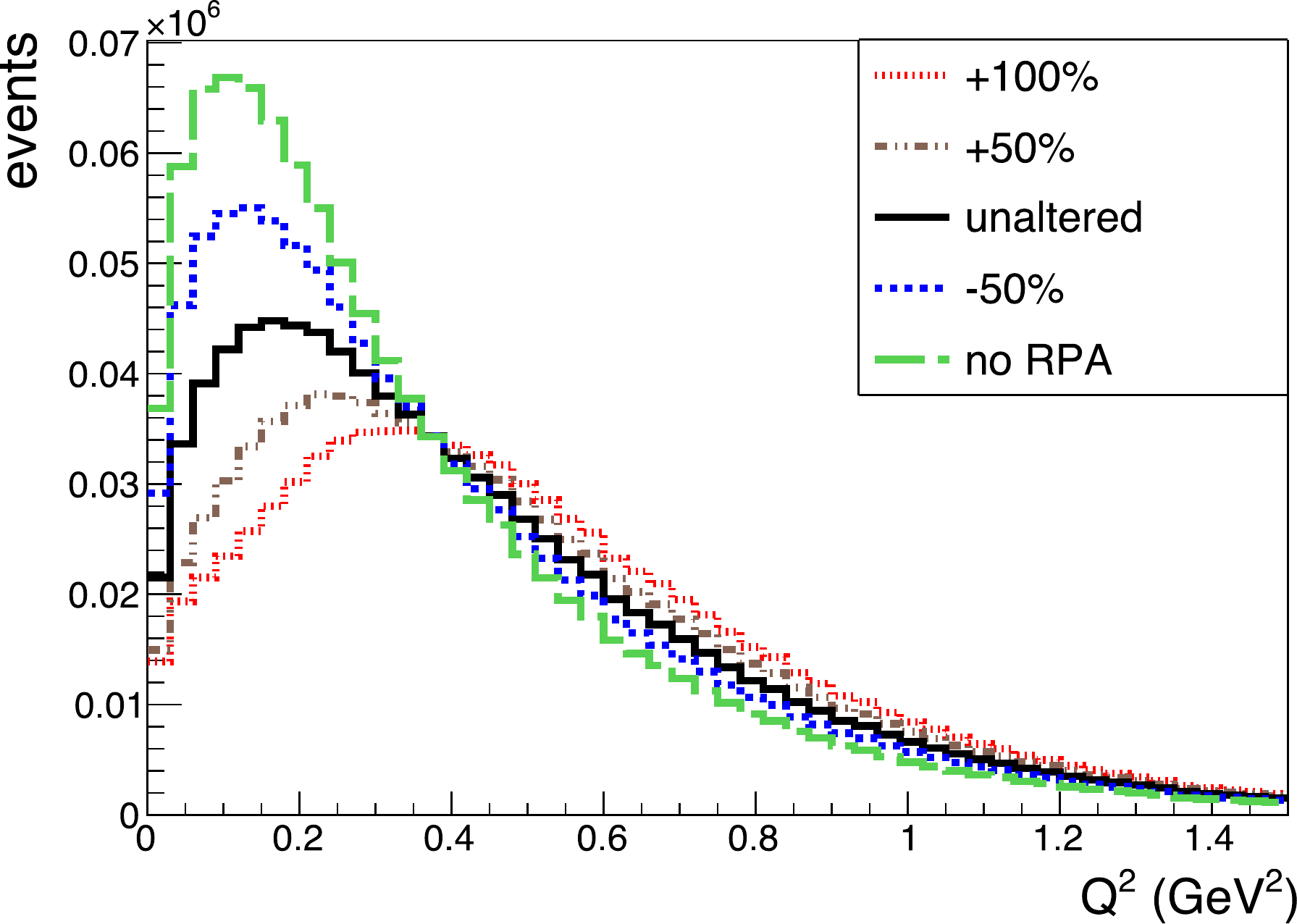}
    \caption{GENIE v3.0.6 prediction of the true $Q^2$ distribution for
$\nu_\mu$ CCQE scattering on \isotope[40]{Ar} in MicroBooNE. The black
histogram shows the untuned result for the G$18\_10a\allowbreak \_02\allowbreak
\_11a$ model set, which includes Valencia RPA corrections to the CCQE cross
section. The other histograms show alterations to the strength of the untuned
CCQE RPA correction calculated via reweighting.}
    \label{fig:truth:compareRPA}
\end{figure}

\textbf{CCQE RPA:} GENIE v3.0.6 G18\_10a\_02\_11a uses the Valencia RPA (Random
Phase Approximation) calculation to correct the CCQE cross section for
nucleon-nucleon long range correlations. This manifests itself predominantly as
a suppression of the CCQE cross section at low $Q^2$. There exists a large body
of evidence in support of such a suppression, but calculations differ on its
exact strength (including theoretical approximations used in the Valencia
prediction), which is not currently well-constrained by data. A new GENIE
reweighting parameter has been developed for this work that extrapolates
linearly from the nominal RPA model (Valencia) to a model in which the RPA
correction is turned off completely. Moving the parameter's value in the
opposite direction strengthens the RPA suppression.
Figure~\ref{fig:truth:compareRPA} illustrates the effect of applying the
parameter to a sample of simulated $\nu_\mu$ CCQE events generated using a
\isotope[40]{Ar} target and the MicroBooNE flux. The low-$Q^2$ portion of the
event distribution shows the expected suppression in the nominal model (black)
relative to the ``RPA correction off'' model (green). The other histograms show
further reweighting-based variations of the RPA strength which follow the
expected trend.

\begin{figure}[hb]
    \centering
    \includegraphics[width=0.45\textwidth]{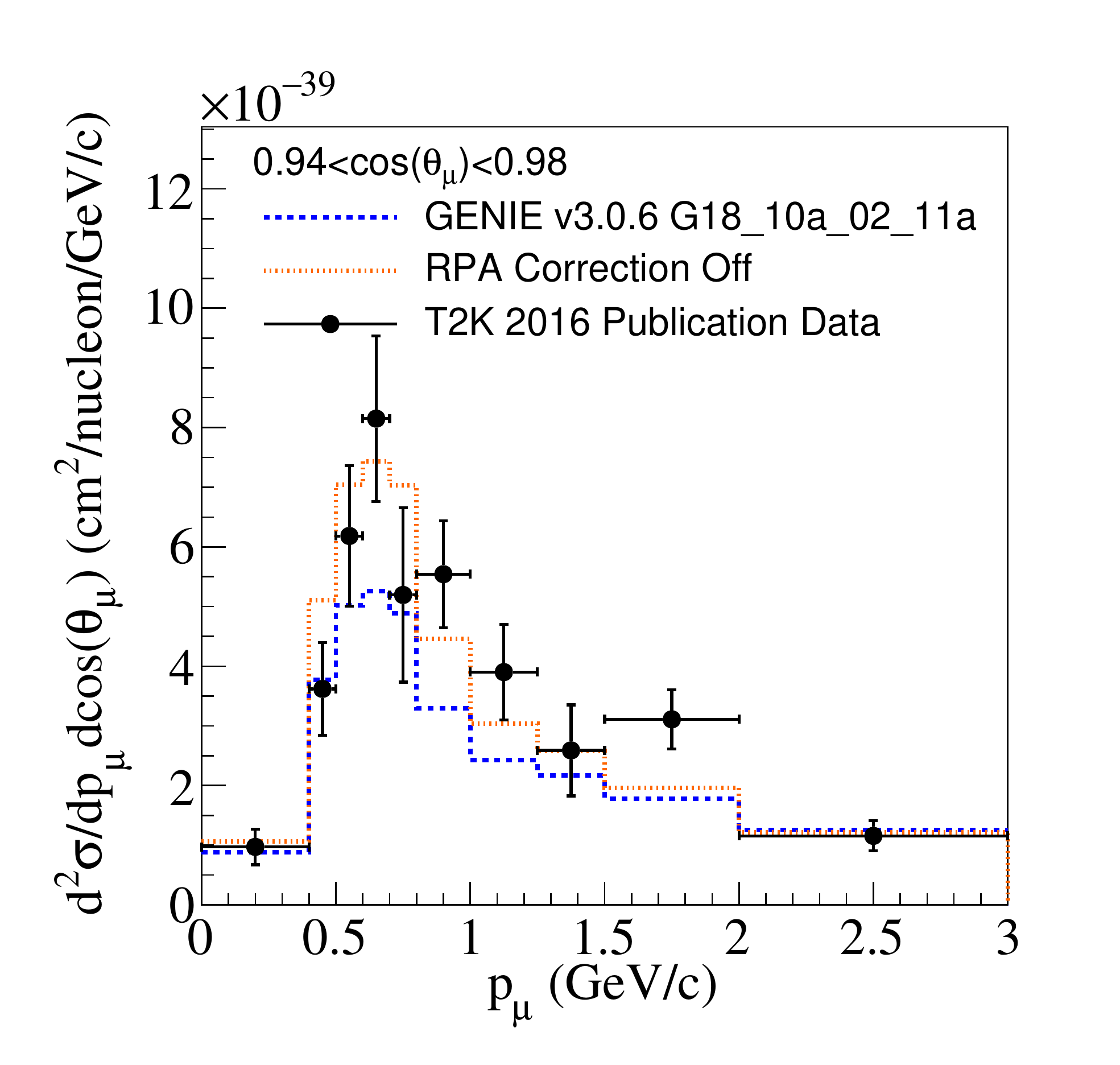}
    \includegraphics[width=0.45\textwidth]{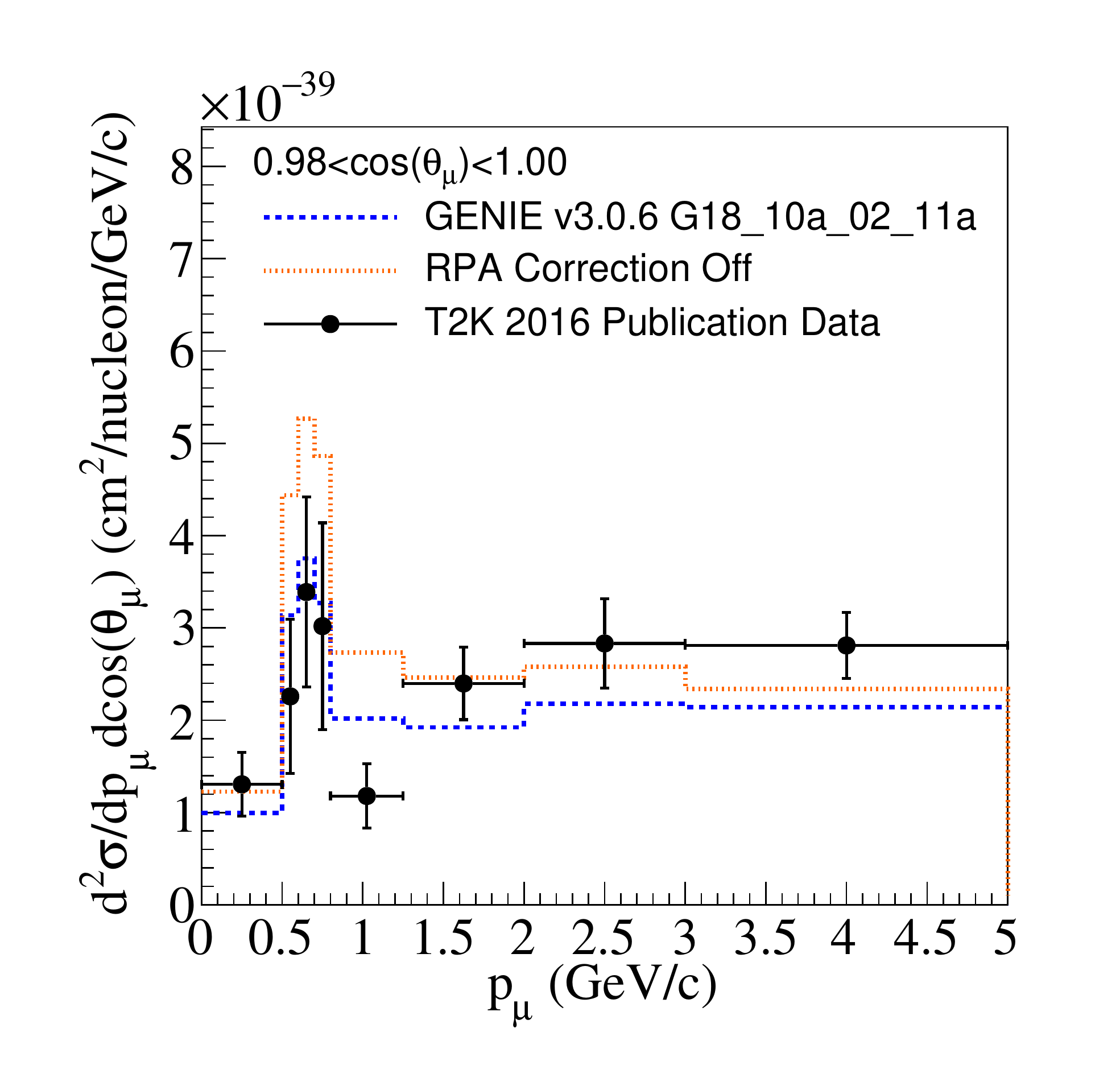}
    \caption{T2K flux-averaged CC0$\pi$ analysis 1 data as a function
of muon momentum for 0.94$<$cos($\theta_{\mu}$)$<$0.98 (top) and
0.98$<$cos($\theta_{\mu}$)$<$1.00 (bottom) compared to the nominal GENIE v3.0.6
G$18\_10a\allowbreak \_02\allowbreak \_11a$ prediction and the prediction with
no CCQE RPA corrections. Both plots are zoomed in on a lower range of muon momentum for readability, although all bins are included in the $\chi_{diag}^2$ calculations in Table~\ref{tab:t2kfits:compareRPA}.}
    \label{fig:t2kfits:compareRPA}
\end{figure}

Figure~\ref{fig:t2kfits:compareRPA} and Table~\ref{tab:t2kfits:compareRPA} show
the impact of RPA corrections on predictions for CC0$\pi$ events at forward
muon scattering angles (low $Q^2$) in T2K, changing between the untuned
prediction (100\% of the nominal RPA strength) and the ``RPA correction off''
prediction.
The effect on the T2K distribution is significant and consistent with what we
would expect; the overall effect of including RPA corrections is to improve the
fit in one angle bin and make it worse in the other bin.  Because of the large
impact on the CCQE cross section and thus on the T2K prediction, we include
this RPA parameter in the fits.


\begin{table}[hbt]
    \centering
        \caption{$\chi_{diag}^2$ for GENIE v3.0.6 G$18\_10a\allowbreak \_02\allowbreak
\_11a$ with and without the Valencia CCQE RPA corrections for T2K CC0$\pi$ data
as shown in Fig.~\ref{fig:t2kfits:compareRPA}. Only the diagonal terms in the
covariance matrix are used in the $\chi_{diag}^2$ calculation. }

\begin{tabular}{|c|c|c|} \hline
$\chi_{diag}^2/\rm{N_{bins}}$     &  0.94$<$cos($\theta_{\mu}$)$<$0.98 & 0.98$<$cos($\theta_{\mu}$)$<$1.00 \\ \hline
RPA Correction & 23.0/11 & 16.2/9 \\ \hline
No RPA Correction & 14.8/11 & 35.9/9 \\ \hline
\end{tabular}
    \label{tab:t2kfits:compareRPA}
\end{table}

\textbf{CC2p2h Normalization:} This parameter changes the overall absolute
normalization of the CC2p2h cross section. A parameter value of 1 corresponds
to the default normalization of CC2p2h interactions in GENIE v3.0.6
G18\_10a\_02\_11a.  A larger value increases the average cross section with no
change in shape.

\begin{figure}[hbt]
    \centering
          \includegraphics[width=0.49\textwidth]{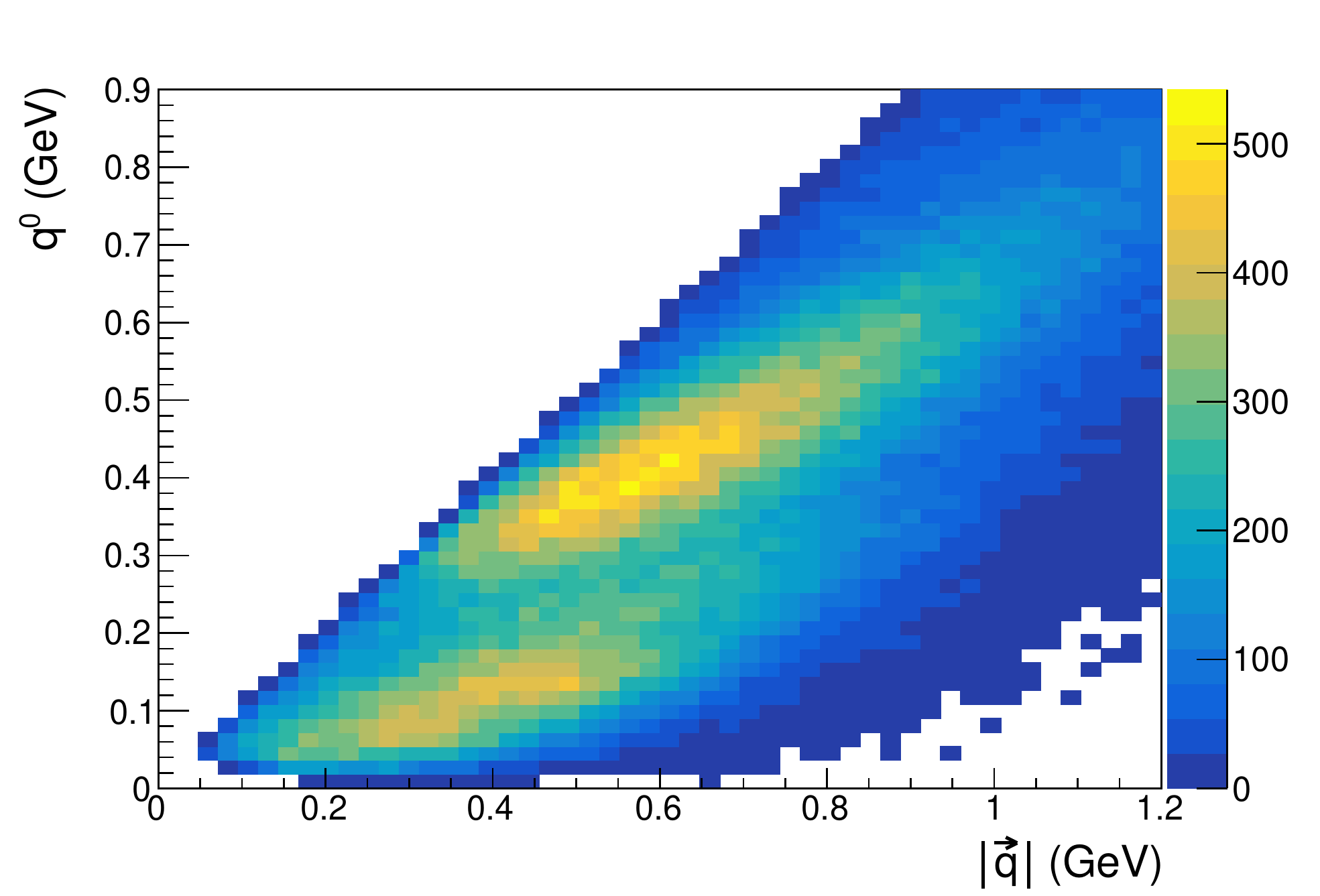} \hfill \includegraphics[width=0.49\textwidth]{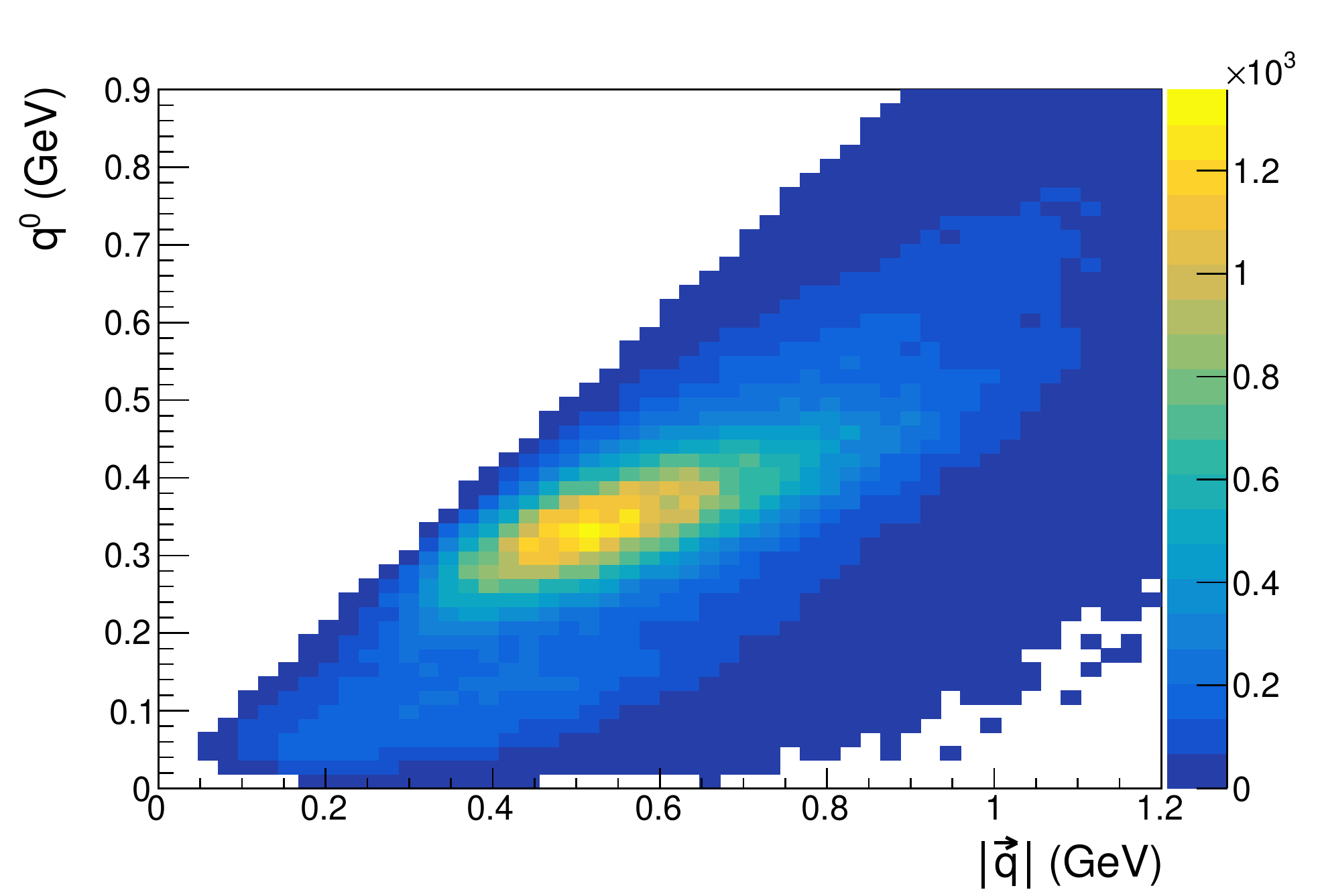} \hfill 
          \includegraphics[width=0.49\textwidth]{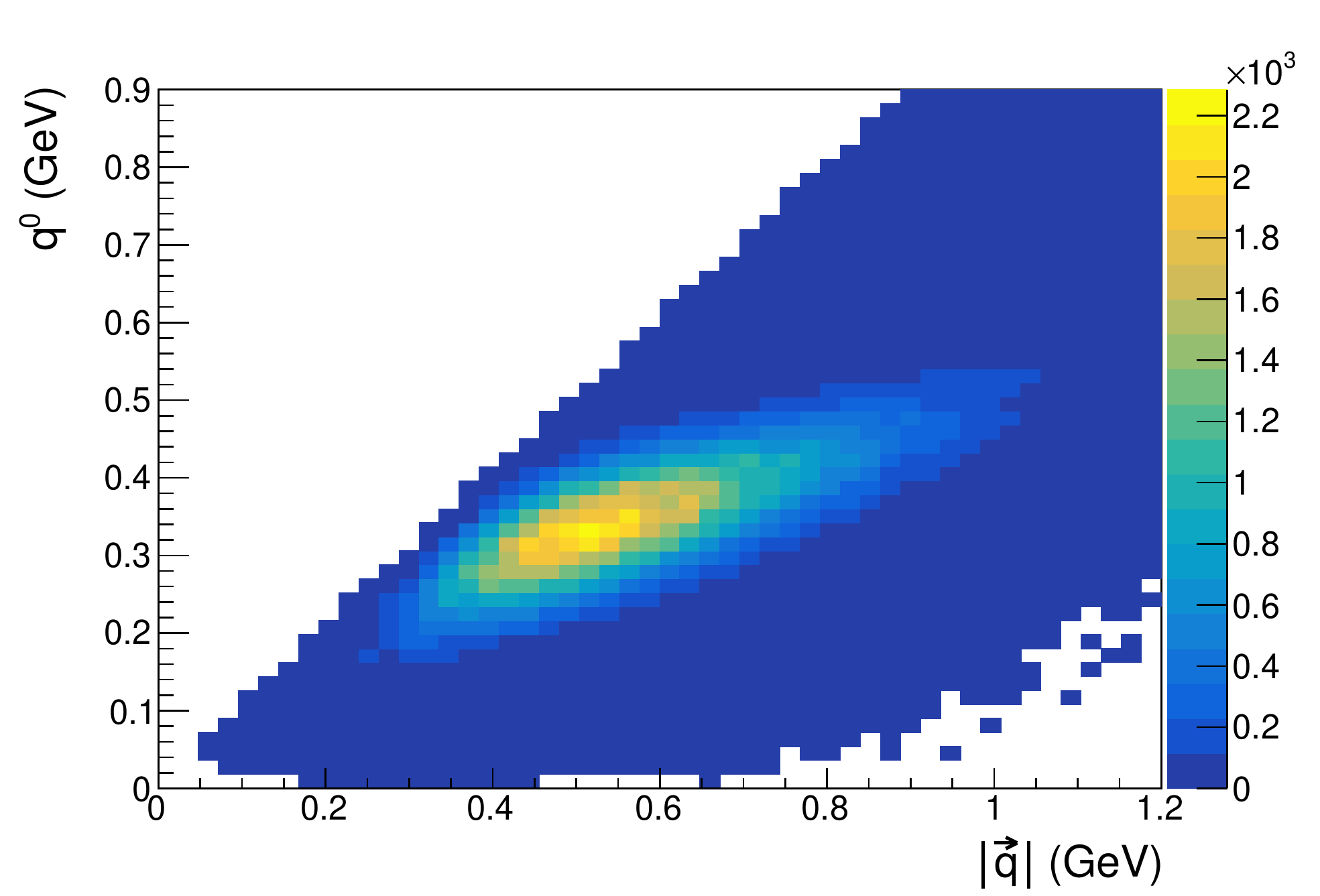}
    \caption{Joint distribution (with arbitrary normalization) of the energy
transfer ($q^0$) and magnitude of the three-momentum transfer ($|\vec{q}|$) for
simulated $\nu_\mu$ CC2p2h events on \isotope[40]{Ar} in MicroBooNE. Top:
Valencia model prediction (CC2p2h shape dial at 0). Middle: Intermediate
prediction (CC2p2h Shape dial at 0.5). Bottom: GENIE empirical model prediction
(CC2p2h shape dial at 1).}
    \label{fig:mecshape:truth}
\end{figure}

\textbf{CC2p2h Shape:} At present, there are substantial differences between
alternative theoretical predictions of lepton kinematics on CC2p2h scattering.
To account for this, a parameter developed for this work changes the shape of
the inclusive CC2p2h differential cross section between the Valencia
calculation~\cite{Gran:2013kda} (the nominal model in our version of GENIE
v3.0.6 G18\_10a\_02\_11a, dial value = 0) and the GENIE Empirical
model~\cite{TeppeiMEC} (an alternative available in other GENIE configurations,
dial value = 1). A linear interpolation is performed, which allows for
continuous variations of the dial on the interval $[0,1]$. The overall
normalization of the cross section is left unaltered.
Fig.~\ref{fig:mecshape:truth} illustrates the effect of the CC2p2h shape dial
variations on the distribution of the leptonic energy and momentum transfer for
simulated CC2p2h events in MicroBooNE. The two distinct peaks seen for the
Valencia calculation (top plot) are replaced by a single peak (middle plot,
dial value = 0.5) as the distribution is reshaped to resemble the GENIE
Empirical model prediction (bottom plot) more closely. A similar kinematic
distribution with a single peak is predicted by other CC2p2h models, e.g.,
SuSAv2-MEC~\cite{SuSAv2}.

\begin{figure}[hbt]
    \centering
        \includegraphics[width=0.45\textwidth]{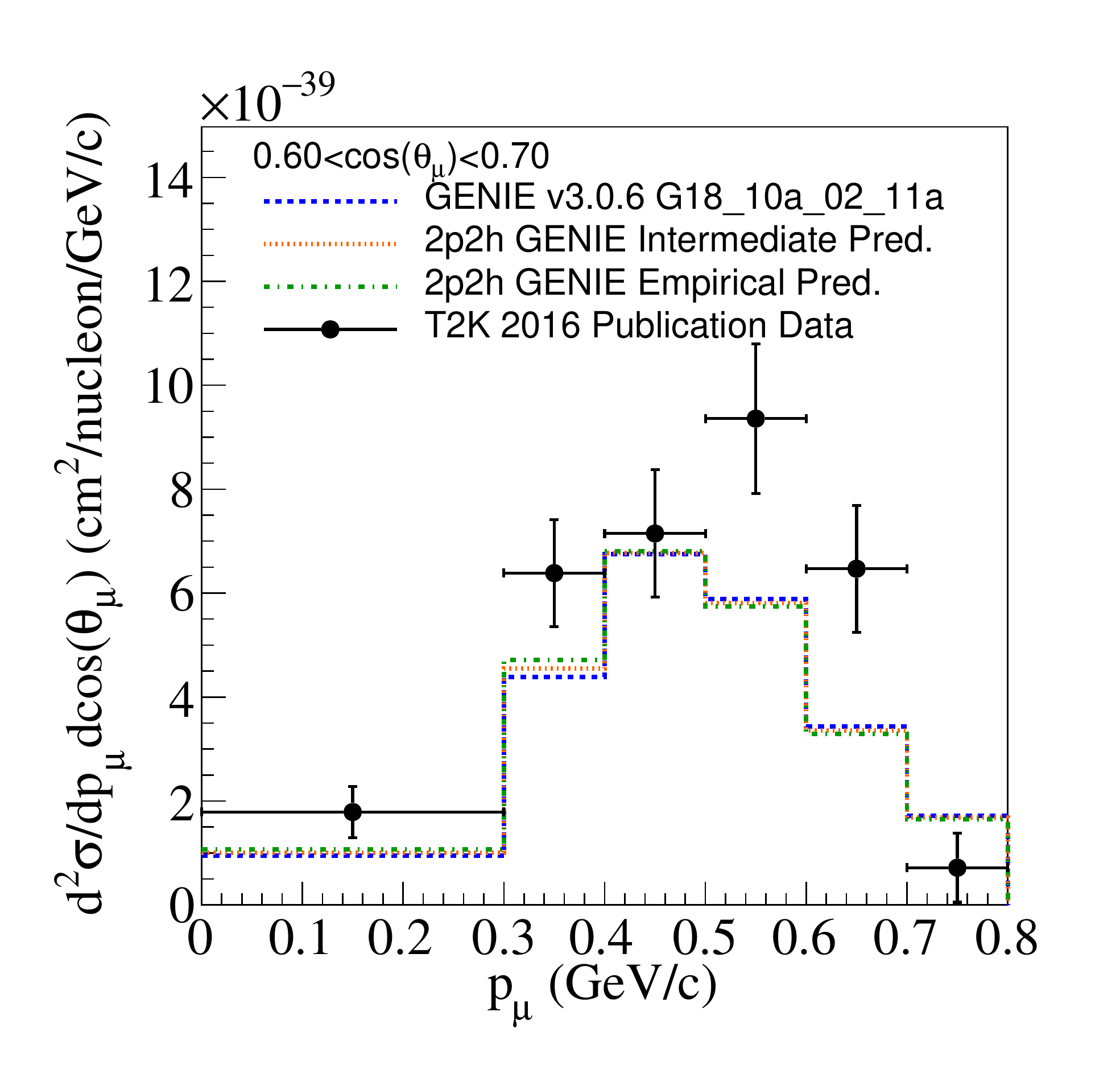}
        \includegraphics[width=0.45\textwidth]{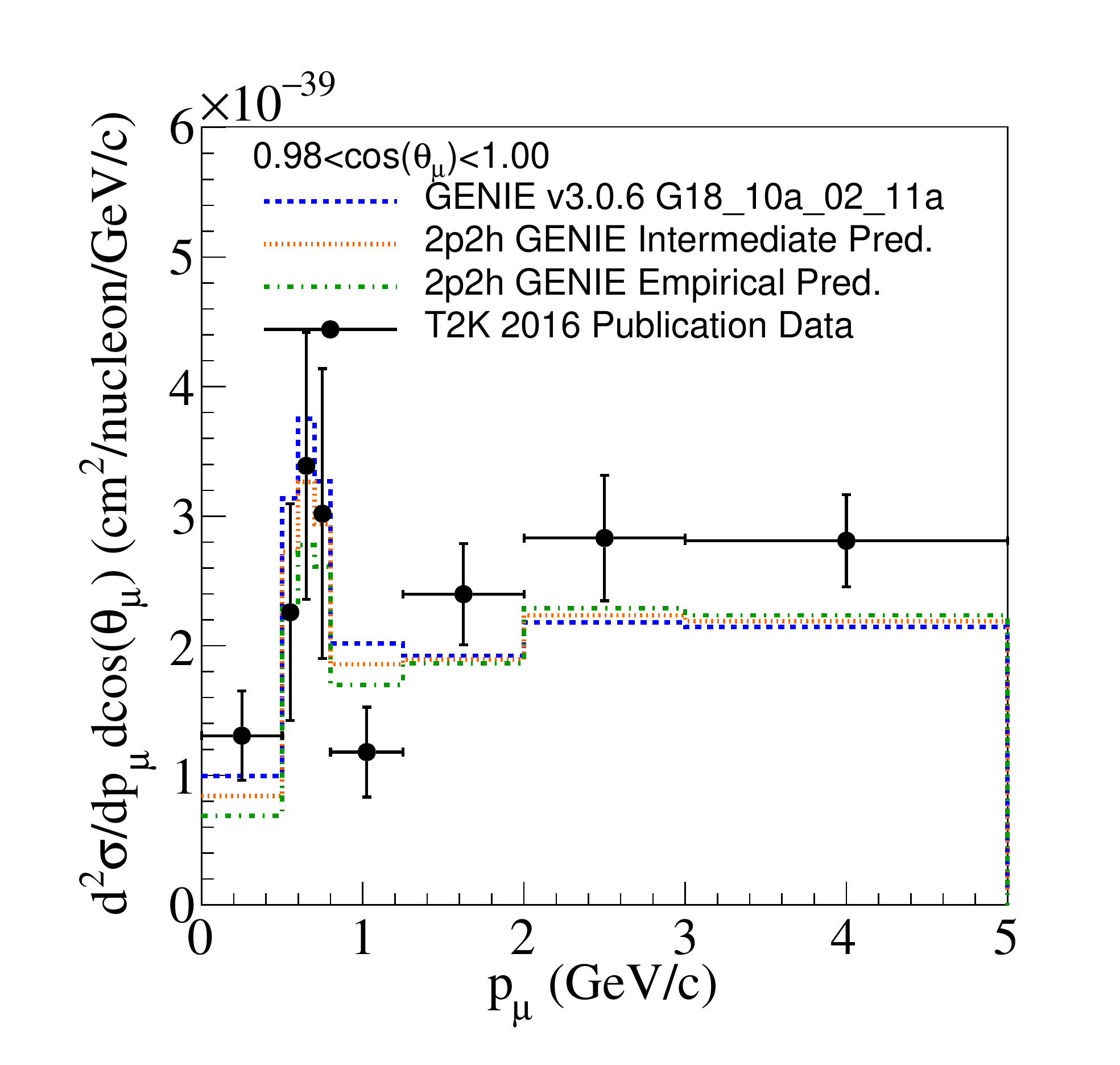}
        \caption{T2K flux-averaged CC0$\pi$ analysis 1 cross-section
measurement as a function of muon momentum for cos($\theta_{\mu}$) between
0.60$<$cos($\theta_{\mu}$)$<$0.70 (top) and 0.98$<$cos($\theta_{\mu}$)$<$1.00
(bottom) compared to the nominal GENIE v3.0.6 G18\_10a\_02\_11a prediction with the Valencia model prediction (CC2p2h Shape dial at 0), the intermediate prediction (CC2p2h Shape dial at 0.5), and the GENIE empirical model prediction (CC2p2h shape dial at 1). Although neither plot shows the overflow bins for better understanding of the differences in performance, all bins are included in the $\chi_{diag}^2$ calculations in Table~\ref{tab:t2kfits:compareMECXSecShape}.}
    \label{fig:t2kfits:compareMECXSecShape}
\end{figure}

Figure~\ref{fig:t2kfits:compareMECXSecShape} shows the effect of changing the
CC2p2h Shape dial from 0 (nominal) to 1 on the predicted CC0$\pi$ cross section
at T2K. Although the CC2p2h shape change is subtle in some regions of phase
space (top plot), it is an important effect in others (bottom plot). The
sensitivity to shape of the muon angle-energy distribution is further
corroborated in the $\chi_{diag}^2$ values reported in
Table~\ref{tab:t2kfits:compareMECXSecShape}. The CC2p2h shape parameter has a significant impact for MicroBooNE.  Since the CC2p2h cross section shape is very much unknown, we include this parameter in our tuning.


\begin{table}[hbt]
    \centering
        \caption{$\chi_{diag}^2$ for GENIE v3.0.6 G$18\_10a\allowbreak \_02\allowbreak
\_11a$ with  the Valencia model prediction (CC2p2h Shape dial at 0), the intermediate prediction (CC2p2h Shape dial at 0.5), and the GENIE empirical model prediction (CC2p2h shape dial at 1) for flux-averaged T2K
CC0$\pi$ analysis 1 data as shown in Fig.~\ref{fig:t2kfits:compareMECXSecShape}. Only the
diagonal terms from the data set covariance matrix are used for the $\chi_{diag}^2$ calculation.
}

\begin{tabular}{|c|c|c|} \hline
$\chi_{diag}^2/\rm{N_{bins}}$     &  0.60$<$cos($\theta_{\mu}$)$<$0.70 & 0.98$<$cos($\theta_{\mu}$)$<$1.00 \\ \hline
Valencia CC2p2h & 21.9/7 & 16.2/9 \\ \hline
Intermediate CC2p2h & 21.2/7 & 13.5/9 \\ \hline
Empirical CC2p2h & 20.6/7 & 12.8/9 \\ \hline
\end{tabular}
    \label{tab:t2kfits:compareMECXSecShape}
\end{table}

The final parameterization for this tuning work consists of four parameters to
be fit: MaCCQE, CC2p2h Normalization, CCQE RPA strength, and CC2p2h shape.

\section{Fit results}
\label{sec:fitresults}

\begin{figure}[htb]
    \centering
    \includegraphics[width=0.45\textwidth]{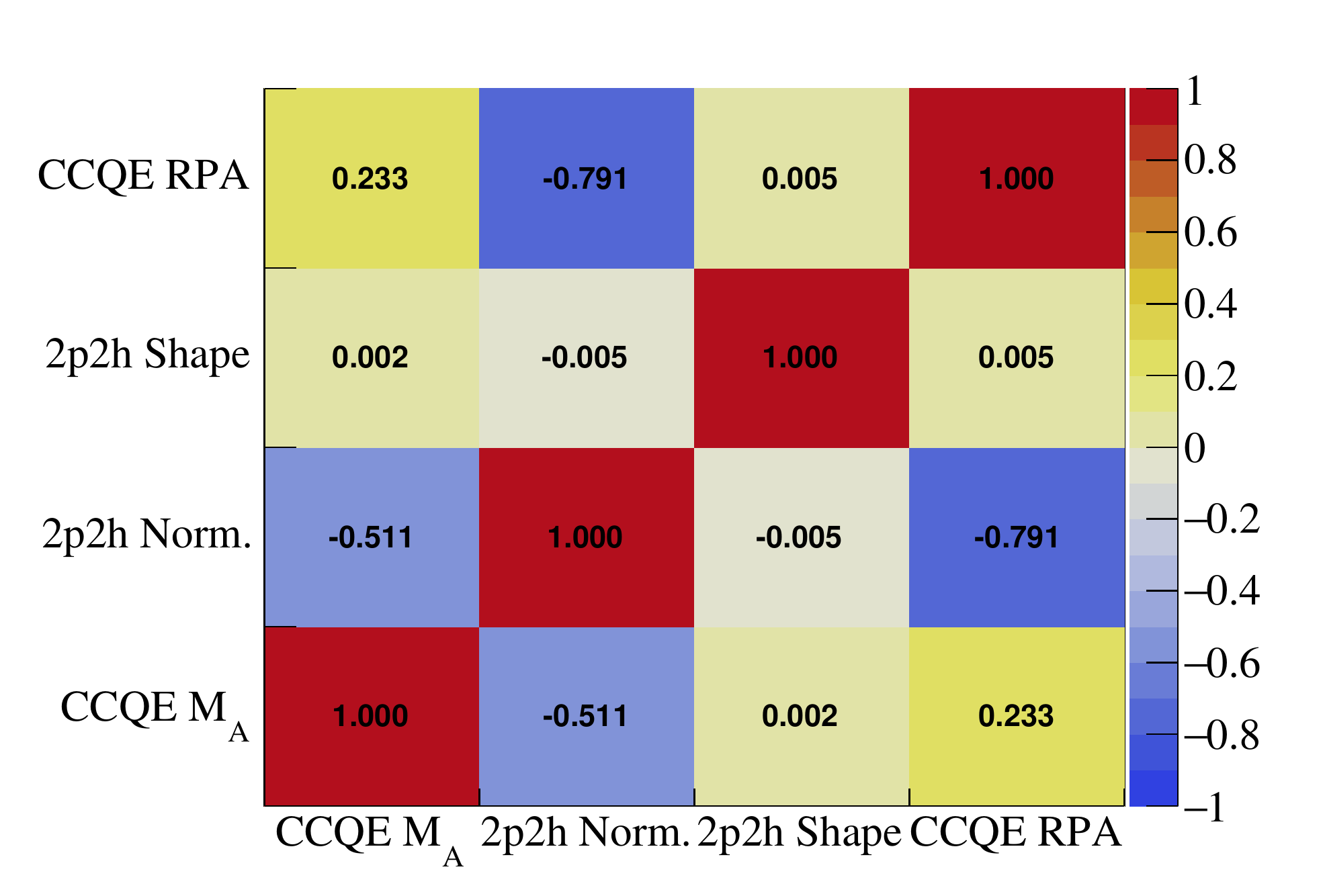}
    \caption{Correlations between parameters after fitting to T2K CC0$\pi$ data.}
    \label{fig:results:corr}
\end{figure}

\begin{figure*}[htb]
    \centering
    \includegraphics[width=\textwidth]{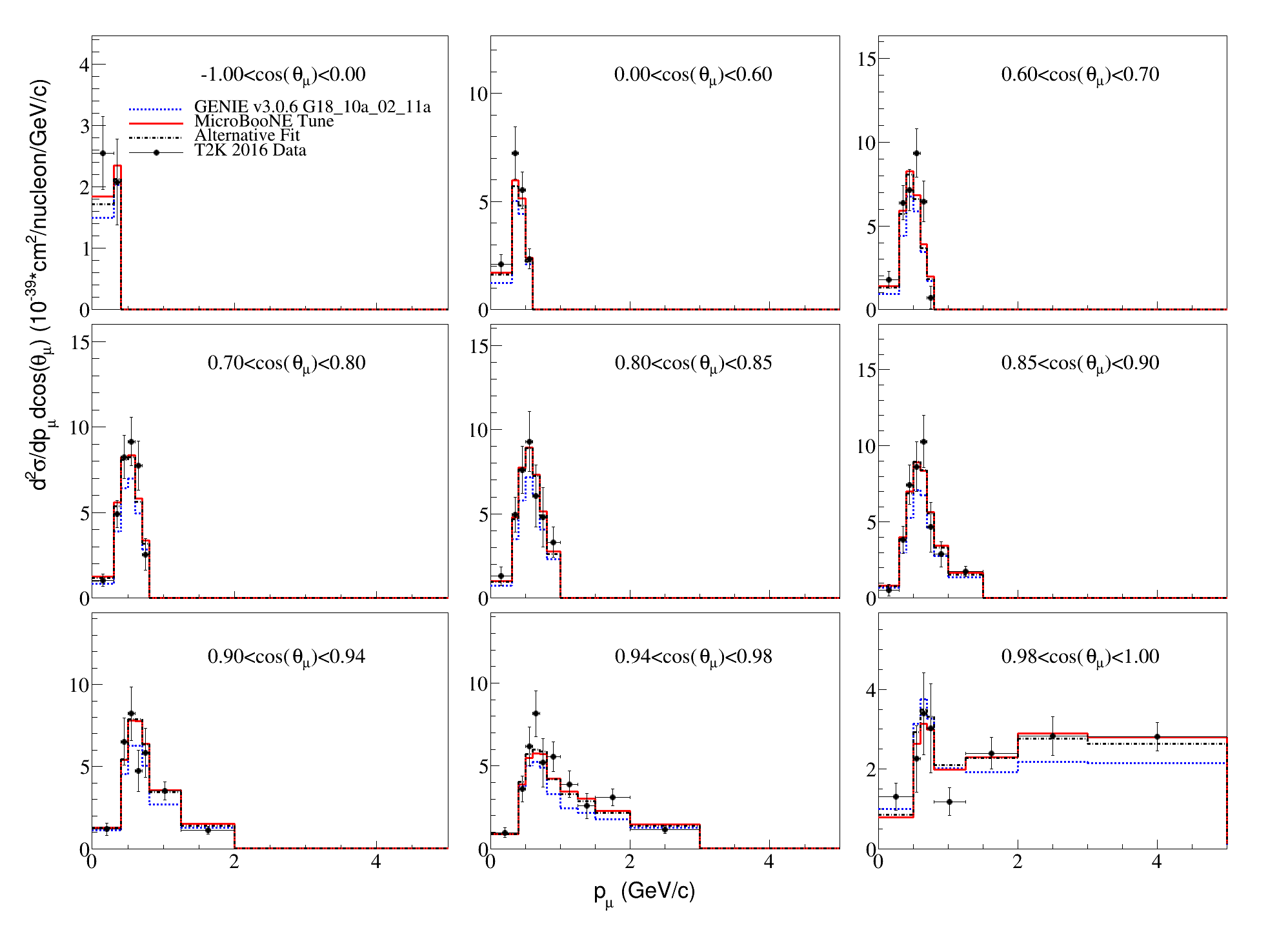}
    \caption{T2K flux-averaged CC0$\pi$ double-differential cross section of lepton
momentum and $\rm{cos(\theta_{\mu})}$ \cite{t2k2016} compared to GENIE v3.0.6 G$18\_10a\allowbreak
\_02\allowbreak \_11a$ ($\chi_{diag}^2/\rm{N_{bins}}$=115.31/67 bins), the ``MicroBooNE Tune''
($\chi_{diag}^2/\rm{N_{bins}}$=63.77/67 bins), and the ``Alternative Fit'' presented in Table~\ref{tab:fit_resultsNormShape} ($\chi_{diag}^2/\rm{N_{bins}}$=63.52/67 bins). Only the diagonal entries in the
covariance matrix are used for the $\chi_{diag}^{2}$ evaluation. While only bins of $\rm{p_{\mu}}<$ 5~GeV are plotted, all bins with data up to $\rm{p_{\mu}}=$ 30~GeV are included in the $\chi_{diag}^2$ calculations.}
    \label{fig:results:fit}
\end{figure*}

We fit four parameters (MaCCQE, CC2p2h Normalization, CCQE RPA Strength, and
CC2p2h Shape) to the T2K CC0$\pi$ cross, section data, neglecting off-diagonal
terms in the T2K data covariance matrix and the highest muon momentum bins (as
described in Sec.~\ref{sec:t2kdata}). Table~\ref{tab:fit_results} shows the
results of three fits, adding in a parameter each time, such that the final row
on the table shows the result of the full four-parameter fit (referred to in
this article as the ``MicroBooNE Tune''). The fitted parameter values in this
line are one of the primary results of this analysis.  Post-fit correlations
between the parameters are shown in Fig.~\ref{fig:results:corr}, and the tuned
model is compared to the T2K data in Fig.~\ref{fig:results:fit}.  Agreement
between data and simulation is measured as a $\chi_{diag}^2$ value across the whole
data set.
As a result of the fit, the total $\chi_{diag}^2/\rm{N_{bins}}$ is reduced from 115.3/67 to 63.8/67
(almost a factor of two) for the data set using the errors from the diagonal of the covariance matrix. Using the full covariance matrix, the $\chi_{full}^2/\rm{N_{bins}}$ is 155.2/67 for the ``MicroBooNE Tune'' and 144.4/67 for the GENIE v3.0.6 G18\_10a\_02\_11a prediction.

The fit results show increases in both MaCCQE (which in large part increases
the CCQE cross section) and the CC2p2h cross-section normalization. It favors a
slight decrease in CCQE RPA strength (85\% of nominal), albeit with a value
consistent within a 1$\sigma$ uncertainty of the Valencia prediction (100\%).
Interestingly, the fit prefers a CC2p2h shape in lepton kinematics that matches
the Empirical CC2p2h model over the Valencia prediction, although the fit
uncertainty is close to the entire range of the parameter, indicating that the
preference is not strong.

In Fig.~\ref{fig:results:corr}, we find fairly strong anti-correlations between
\mbox{MaCCQE} and CC2p2h normalization, because increasing MaCCQE increases the
CCQE cross-section normalization (with some additional changes in shape).
Strong anti-correlations are also seen between the CC2p2h normalization and RPA
strength parameters, which is to be expected because both have importance at
forward muon angles. These anti-correlations suggest ambiguities in the tuning
procedure, where different fitted parameters can have similar effects on the
prediction. Therefore, one cannot assume that the central-value result of this
tune has the correct ratio of contributions from the CCQE and CC2p2h processes.
However, this tune does successfully adjust the overall rate and shape of the
prediction in lepton kinematics for CC0$\pi$ interactions. Ambiguities between
individual interaction processes can be correctly accounted for if the
correlation matrix is taken into account in the treatment of uncertainties. The
correlations between the CC2p2h normalization and shape parameters are small,
consistent with the design of these parameters to be orthogonal.

Tests of the robustness of the fit are important.  The variation of parameters
in Table~\ref{tab:fit_results}  gives some indication of how choices of
parameters impact the fit results.  The variation of fit results as the number
of parameters increases is not significant.  The best-fit values for MaCCQE and
CC2p2h normalization are fairly constant within uncertainties, with and without
the shape parameters.  The fitted value of CC2p2h normalization changes
depending on the fitted value of RPA strength, which is consistent with the
large anti-correlations between these parameters seen in the four-parameter fit
covariance matrix.

As another test of robustness, the recent method of Koch~\cite{kochChi2} was used to include the effects of correlations between data bins.  This method empirically separates the correlations into normalization (mostly due to uncertainties in neutrino flux calculations for neutrino experiments) and shape components when calculating the chi-squared ($\chi^2_{Koch}$).  As a result, Peelle's Pertinent Puzzle can be mitigated. The best-fit parameter values obtained using this method are within the uncertainties provided by MINUIT of the values reported in Table~\ref{tab:fit_resultsNormShape}, with a difference of less than 10\% in both $\chi^2_{Koch}$ and $\chi^2_{diag}$ with respect to the T2K data. Figure~\ref{fig:results:fit} shows only small differences bin-to-bin visually between the MicroBooNE Tune and this ``Alternative fit.'' This gives evidence that the fitting method was reasonable.

In addition, different parameters were chosen (e.g. CCQE
normalization vs. MaCCQE and different formulations of the shape parameters)
with minimal change in the resulting agreement between data and simulation. In addition, alternate fits with different
starting values, both small and random, give almost identical fit values and
uncertainties. This gives confidence that the choice of fit parameters was
robust.


\begin{table*}[tb]

    \renewcommand*{\arraystretch}{1.5}
        \caption{Tuned parameter values and uncertainties after fitting to T2K
CC0$\pi$ data for the nominal simulation and three tunes that build to the
final four parameter tune. Note that post-fit $\chi^2$ values are quoted here
only for the 58 bins included in the fit (excluding the highest muon momentum
bin in each cos $\theta$ bin), and using diagonal elements of the covariance matrix only. In the text and figures, pre- and post-fit
$\chi^2$ comparisons are also quoted for the full T2K data set of 67 bins.}
    \centering
    \begin{tabular}{|L{3.5cm}|L{2.5cm}|L{2.4cm}|L{3.3cm}|L{2.4cm}||L{2cm}|}
        \hline
         & \raggedright MaCCQE fitted value & \raggedright CC2p2h Norm. fitted
value & \raggedright CCQE RPA Strength fitted value & \raggedright CC2p2h Shape
fitted value & T2K $\chi_{diag}^2/\mathrm{N}_{bins}$ \\
         \hline
        \raggedright Nominal (untuned) & 0.961242 GeV & 1 & 100\% & 0 & 106.7/58 \\
        \raggedright Fit MaCCQE + CC2p2h Norm. & 1.14$\pm$0.07 GeV & 1.61$\pm$0.19 & 100\% (fixed) & 0 (fixed) & 71.8/58\\
        \raggedright Fit MaCCQE + CC2p2h Norm. + CCQE RPA Strength & 1.18$\pm$0.08 GeV & 1.12$\pm$0.38 & (64$\pm$23)\% & 0 (fixed) & 69.7/58 \\
        \raggedright Fit MaCCQE + CC2p2h Norm. + CCQE RPA Strength + CC2p2h Shape & 1.10$\pm$0.07 GeV & 1.66$\pm$0.19 & (85$\pm$20)\% & 1$^{+0}_{-0.74}$ & 52.5/58 \\
        \hline
    \end{tabular}

    \label{tab:fit_results}
\end{table*}

\begin{table*}[tb]

    \renewcommand*{\arraystretch}{1.5}
        \caption{Parameter values and uncertainties from nominal GENIE v3.0.6, the ``MicroBooNE Tune,'' and the ``alternate fit.'' Post-fit $\chi^2$ values are quoted
only for the 58 bins included in the fit (excluding the highest muon momentum
bin in each cos $\theta$ bin) using the diagonal elements of the covariance matrix only ($\chi^2_{diag}$), the Koch norm-shape covariance matrix~\cite{kochChi2} ($\chi^2_{Koch}$), and the full covariance matrix ($\chi^2_{full}$). Note that $\chi^2_{diag}$ is the figure-of-merit that is minimized in the ``MicroBooNE Tune'' fit.} 
    \centering
    \begin{tabular}{|L{3.2cm}|L{1.7cm}|L{1.7cm}|L{1.7cm}|L{2.4cm}||L{1.7cm}|L{1.7cm}|L{1.7cm}|}
        \hline
         & \raggedright MaCCQE fitted value & \raggedright CC2p2h Norm. fitted
value & \raggedright CCQE RPA Strength fitted value & \raggedright CC2p2h Shape
fitted value & T2K $\chi_{diag}^2/\mathrm{N}_{bins}$ & T2K $\chi_{Koch}^2/\mathrm{N}_{bins}$  & T2K $\chi_{full}^2/\mathrm{N}_{bins}$  \\
         \hline
        \raggedright Nominal (untuned) & 0.961242 GeV & 1 & 100\% & 0 & 106.7/58 & 149.83/58  & 97.56/58  \\
        \raggedright ``MicroBooNE Tune'' & 1.10$\pm$0.07 GeV & 1.66$\pm$0.19 & (85$\pm$20)\% & 1$^{+0}_{-0.74}$ & 52.5/58 & 110.58/58 & 103.84/58 \\
        \raggedright ``Alternate fit'' & 1.04$\pm$0.10 GeV & 1.44$\pm$0.42 & (67$\pm$16)\% & 0.91$^{+0.09}_{-0.18}$ & 55.51/58 & 100.59/58 & 91.68/58  \\ \hline 
    \end{tabular}

    \label{tab:fit_resultsNormShape}
\end{table*}


The uncertainties shown in Table~\ref{tab:fit_results} are the post-fit
uncertainties given by MINUIT.
MicroBooNE analyses adopt uncertainties that cover the results of all three
fits presented in Table~\ref{tab:fit_results}: an uncertainty
of 0.1 GeV on the parameter MaCCQE, 0.5 on the CC2p2h normalization, and 40\% on
the CCQE RPA strength. Because we expect MicroBooNE data to have a better ability to distinguish between 2p2h models than the T2K data, we adopt an uncertainty that covers the full allowed range of the parameter for CC2p2h Shape: $1^{+0}_{-1}$.

\FloatBarrier
\subsection{``MicroBooNE Tune'' Comparison to MicroBooNE Data}
\label{sec:fitresults:uboone}

Because the aim of this tuning work is to support MicroBooNE analyses, it is
important to compare the ``MicroBooNE Tune'' to MicroBooNE data. While the T2K
data are in a similar energy range to MicroBooNE, they are on a different
nuclear target. Therefore it is imperative to check that the fitted result within
uncertainties can predict MicroBooNE's measured argon-target interactions.
Comparisons of the tuned and untuned GENIE v3 models to MicroBooNE data are
provided in this section for generic neutrino scattering, $\nu_\mu$ CC
inclusive events, and exclusive one-muon, one-proton ($1\mu1p$) final states
consistent with CCQE kinematics.
The goal is to have meaningful comparisons, but no attempt is made to be comprehensive. As is the case for any neutrino interaction model, the suitability of the ``MicroBooNE Tune'' (and its associated uncertainties) for any specific analysis must be determined on a case-by-case basis. Further data-driven model constraints will often be essential for achieving sufficient precision.
However, based on the overall improvement seen in the description of MicroBooNE
data across many event selections and observables, the collaboration has
adopted the ``MicroBooNE Tune'' described herein as the base neutrino interaction
model for all current analyses, including those investigating the MiniBooNE
LEE~\cite{PRLeLEE,WCeLEE,PeLEE,DLeLEE,gLEE} and neutrino-argon cross
sections~\cite{NuMInueDifferential,WCCCincl}.

\begin{figure}[htb]
    \centering
    \subfigure[Simulated neutrino interactions predicted by GENIE v3.0.6 G18\_10a\_02\_11a]{%
    \label{fig:results:ubooneWC:notune}%
	\includegraphics[width=0.5\textwidth]{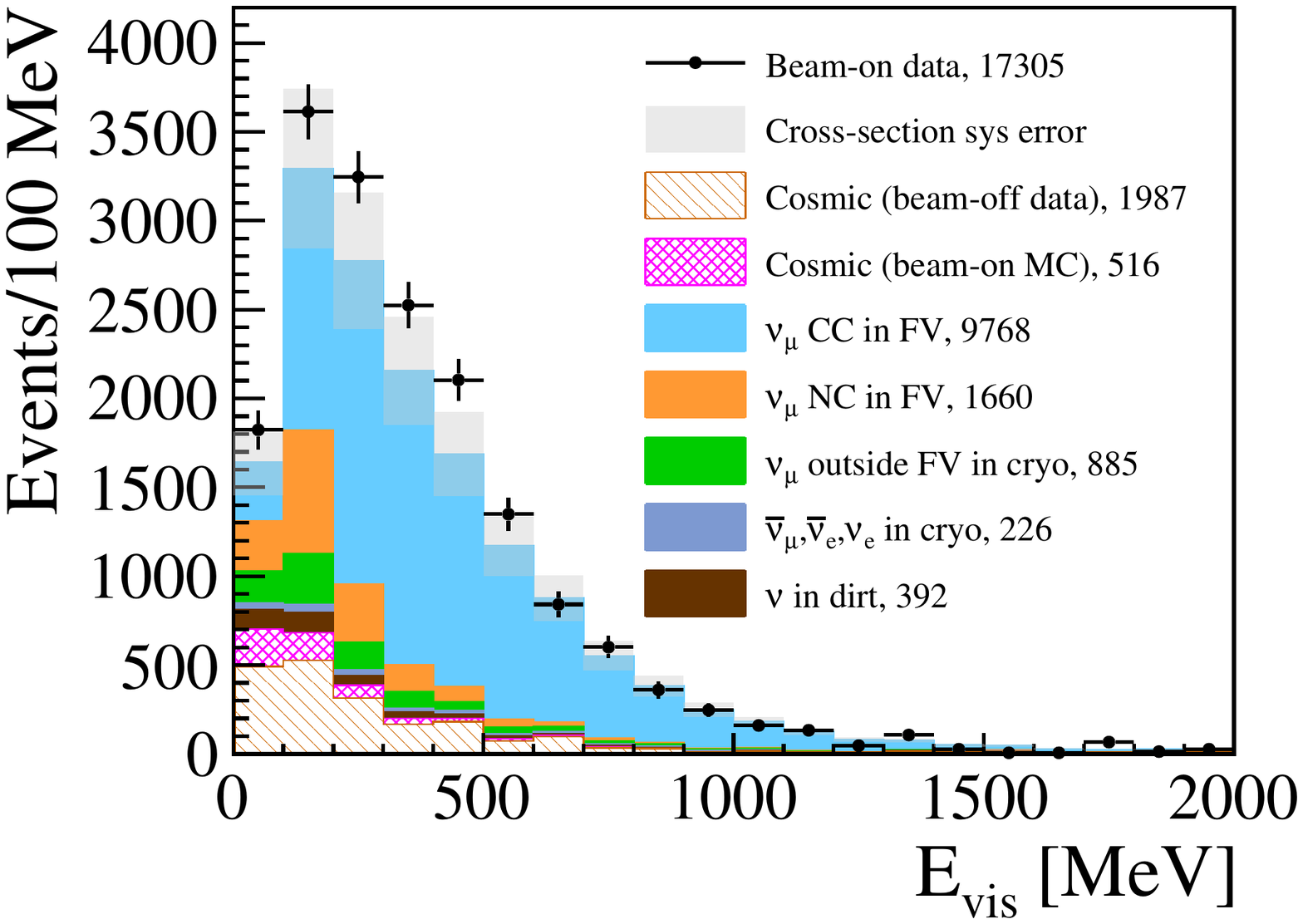}%
	}%
	\hfill
    \subfigure[Simulated neutrino interactions predicted by the ``MicroBooNE
Tune'' applied to GENIE v3.0.6 G18\_10a\_02\_11a]{
    \label{fig:results:ubooneWC:tuned}
	\includegraphics[width=0.5\textwidth]{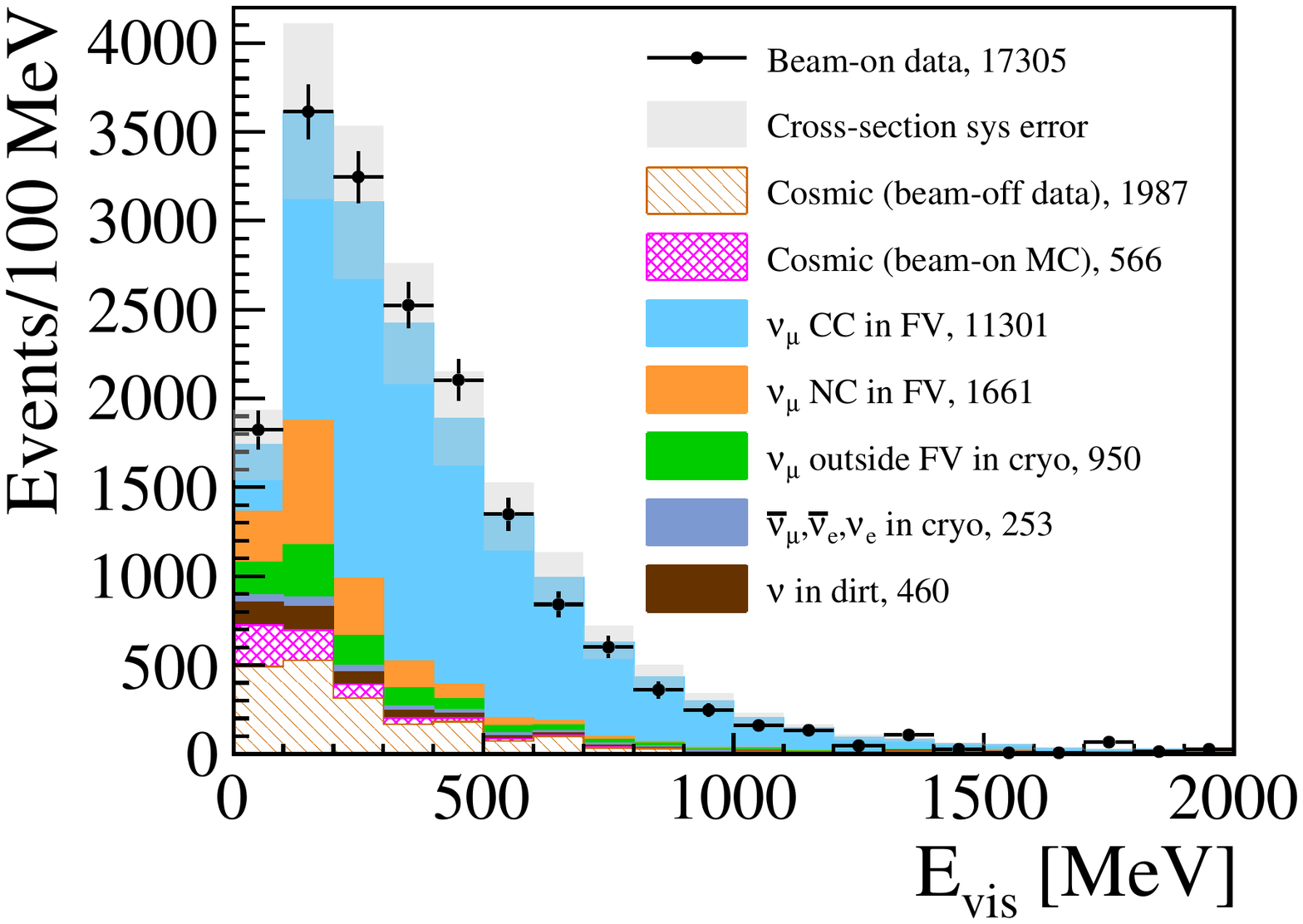}%
	}%
    \caption{Total visible energy of events selected in the MicroBooNE detector
using the generic neutrino detection described in~\cite{ubWCgeneric}, compared
to MicroBooNE simulation before and after the ``MicroBooNE Tune'' has been
applied. The gray area indicates uncertainties on the cross-section
model only (including uncertainties on the tuned parameters, the new
uncertainties presented in Sec.~\ref{sec:par:notfit}, and other uncertainties
as recommended by the GENIE collaboration). The tuned model shows significantly
better agreement with the data.}%
    \label{fig:results:ubooneWC}%
\end{figure}

Figure~\ref{fig:results:ubooneWC} shows the events selected in the MicroBooNE
detector using the generic neutrino detection described in
Ref.~\cite{ubWCgeneric}, plotted as a function of visible energy. The same
selected data events are shown in both panels, but the simulation uses untuned
GENIE v3.0.6 G$18\_10a\allowbreak \_02\allowbreak \_11a$ in
Fig.~\ref{fig:results:ubooneWC:notune} and the simulation  computed with the
``MicroBooNE Tune'' is applied in Fig.~\ref{fig:results:ubooneWC:tuned}. The
tune increases the normalization of the simulation, decreasing the
data/simulation ratio from 1.12 (untuned) to 1.01 (``MicroBooNE Tune'').

\begin{figure*}[htb]
    \centering
  \includegraphics[width=\textwidth, trim=5 5 15 15, clip]{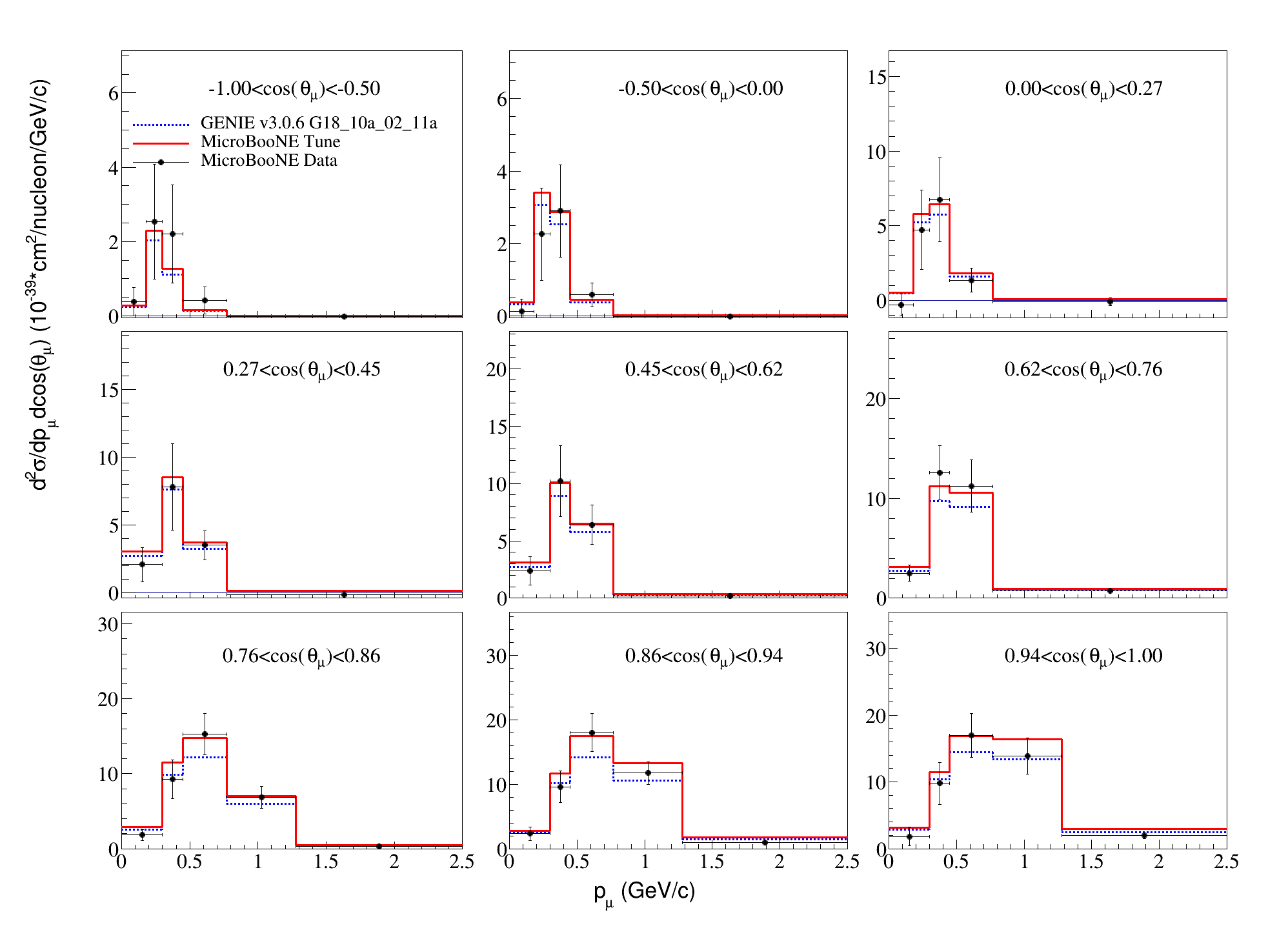}

    \caption{MicroBooNE flux-averaged $\nu_\mu$ CC inclusive
double-differential cross section as a function of muon momentum and
$\rm{cos(\theta_{\mu})}$~\cite{MCC8_CCinc}, compared to GENIE v3.0.6
G18\_10a\_02\_11a ($\chi_{full}^2/\rm{N_{bins}} = $105.41/42) and the ``MicroBooNE Tune'' central value ($\chi_{full}^2/\rm{N_{bins}} = $140.55/42) to T2K CC0$\pi$ data.}
    \label{fig:results:microMCC8}
\end{figure*}

Figure~\ref{fig:results:microMCC8} shows GENIE v3.0.6 G18\_10a\_02\_11a and the
``MicroBooNE Tune'' central value (neglecting uncertainties on the predictions)
compared to the double-differential cross section for CC inclusive interactions
measured in the MicroBooNE detector as a function of lepton momentum and
$\rm{cos(\theta_{\mu})}$~\cite{MCC8_CCinc}. Table~\ref{tab:microMCC8} provides a comparison of $\chi_{full}^2$ values using the full covariance matrix for the complete data set and binned in angle.  As seen elsewhere, a major
effect of the tune is to increase the normalization of the prediction. 
However, the value of $\chi_{full}^2/\rm{N_{bins}}$ in Table~\ref{tab:microMCC8} for the full angular range increases from 105.41/42 (untuned GENIE prediction) to 140.55/42 (``MicroBooNE Tune"). Although the match is poor in both cases, we find that the large $\chi_{full}^2/\rm{N_{bins}}$ value is driven by the highest muon momentum bins for $\rm{cos(\theta_{\mu})}$ approaching 1. For example, the measurement sits below both predictions and has a very small uncertainty in the highest muon momentum bin in the $0.86 \leq \rm{cos(\theta_{\mu})} \leq 0.94$ angular bin. Removing this bin from the comparison gives an overall $\chi_{full}^2/\rm{N_{bins}}$ of 69.7/41 (GENIE v3) or 90.2/41 (``MicroBooNE Tune").  It also reduces the $\chi_{full}^2$ in the $0.86 \leq \rm{cos(\theta_{\mu})} \leq 0.94$ angular bin to 6.2 (GENIE v3) or 8.3 (``MicroBooNE Tune").
We find that the tuning has provided a better description of the data in some regions of phase space, notably at moderate muon production angles and momenta. However, there remains room for improvement in the description at high muon momentum and at very forward-going scattering angles.  The alternative fit using~\cite{kochChi2} described in Sect.~\ref{sec:fitmeth} results in a total $\chi_{full}^2/\rm{N_{bins}}$ of 130.29/42, a 7.2\% improvement over the ``MicroBooNE Tune" value. All the values of $\chi_{full}^2$ are large.  As the $\chi_{full}^2/\rm{N_{bins}}$ differ by less than one unit of $\chi_{full}^2$/bin across all three cases, we conclude that this shows approximately consistent agreement of all three model sets with these data. 

\begin{table*}[htb]

    \renewcommand*{\arraystretch}{1.5}
        \caption{$\chi^2$ for the results of this work in
Fig.~\ref{fig:results:microMCC8} relative to MicroBooNE CC inclusive
double-differential data using the full covariance matrix of the data set for
$\chi_{full}^2$ calculations. Included are both $\chi_{full}^2$ values for the full data set
and for each slice of cos($\theta_{\mu}$). 
The per-slice $\chi^{2}$ values reported here do not include
correlations between different slices.}
    \centering
  
            \begin{tabular}{|c|c|c|}
        \hline
        $\chi_{full}^2/\rm{N_{bins}}$ & GENIE v3.0.6 G$18\_10a\allowbreak \_02\allowbreak \_11a$ & ``MicroBooNE Tune'' \\ \hline
        Full Data Set & 105.41/42 & 140.55/42 \\ \hline
       -1.00$<\rm{cos(\theta_\mu)}<$-0.50 & 3.50/5 bins & 3.47/5 \\ \hline
       -0.50$<\rm{cos(\theta_\mu)}<$0.00 & 2.80/5 bins & 3.07/5 \\ \hline
       0.00$<\rm{cos(\theta_\mu)}<$0.27 & 4.54/5 bins & 4.15/5 \\ \hline
       0.27$<\rm{cos(\theta_\mu)}<$0.45 & 2.95/4 bins & 2.61/4  \\ \hline
       0.45$<\rm{cos(\theta_\mu)}<$0.62 & 2.37/4 bins & 2.09/4 \\ \hline
       0.62$<\rm{cos(\theta_\mu)}<$0.76 & 7.73/4 bins & 8.36/4 \\ \hline
       0.76$<\rm{cos(\theta_\mu)}<$0.86 & 10.89/5 bins &  7.92/5 \\ \hline
       0.86$<\rm{cos(\theta_\mu)}<$0.94 & 34.36/5 bins  & 43.90/5 \\ \hline
       0.94$<\rm{cos(\theta_\mu)}<$1.00 & 7.60/5 bins & 12.41/5 \\ \hline
    \end{tabular}

    \label{tab:microMCC8}
\end{table*}

\begin{figure*}
    \centering
    \begin{overpic}[width=0.329\textwidth]{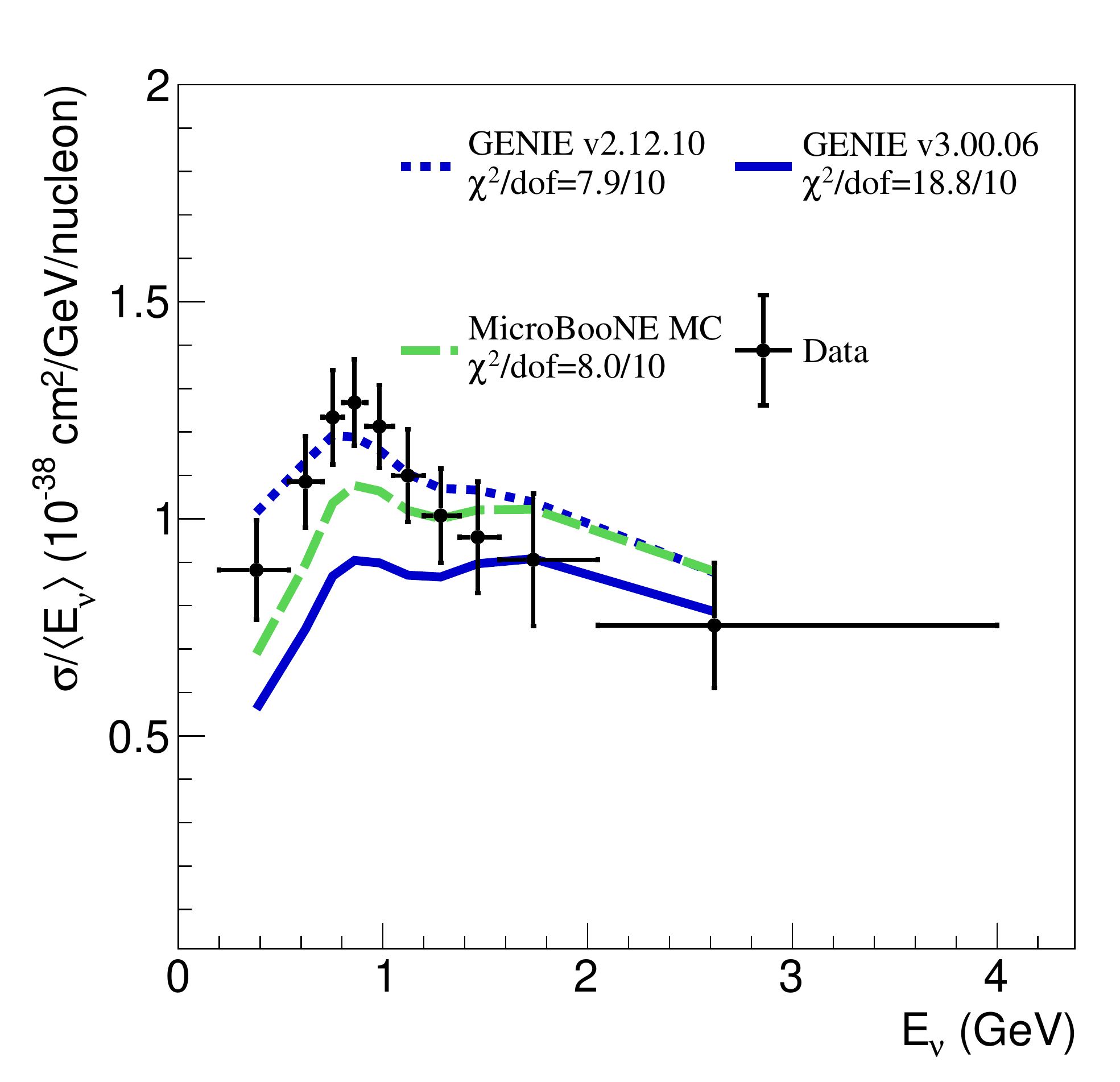}
    \put(50,-3){\footnotesize{(a)}}
    \put(17,91){\textsf{MicroBooNE $\mathsf{5.3\times10^{19}}$ POT}}
   \end{overpic}
   \begin{overpic}[width=0.329\textwidth]{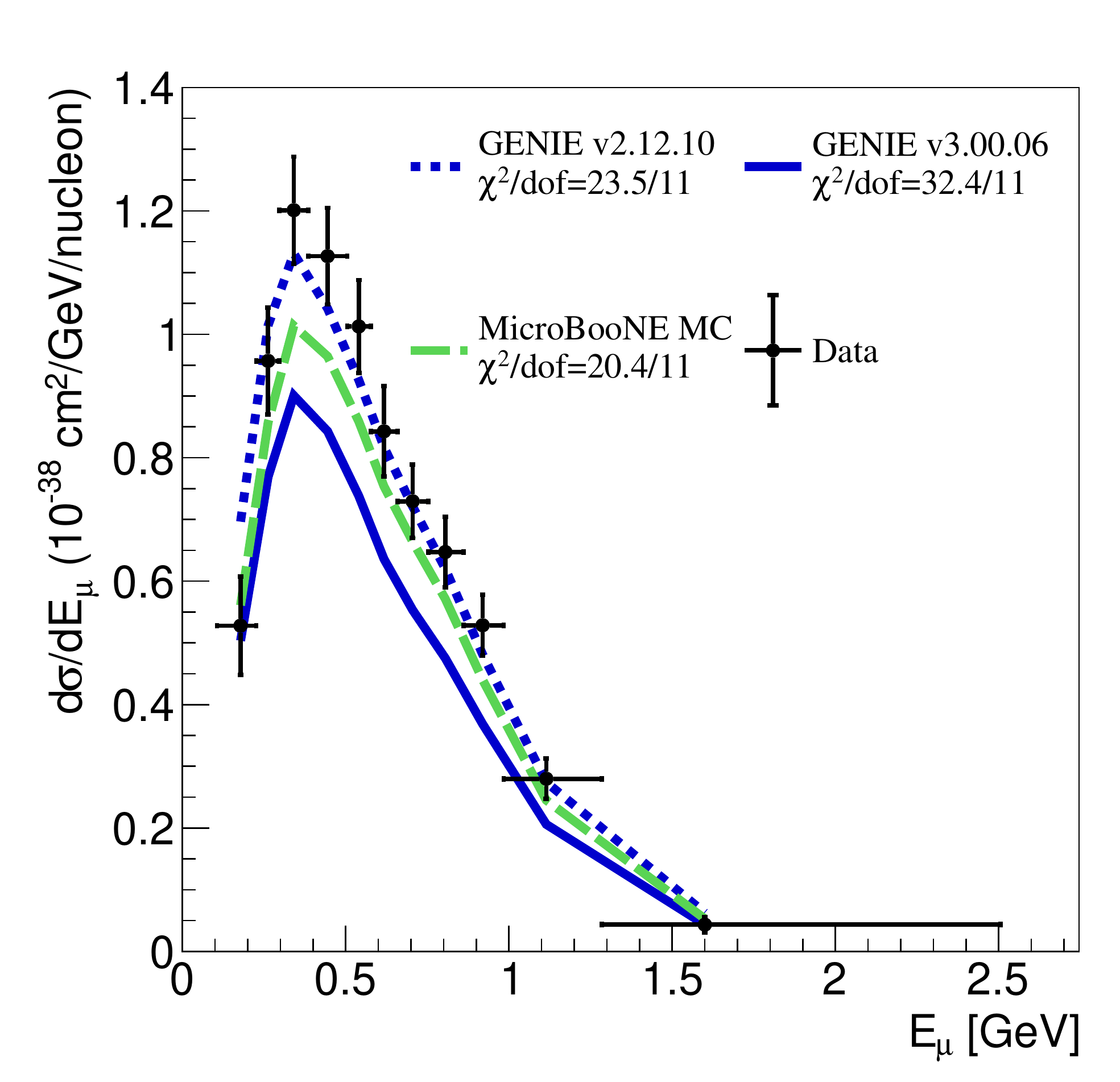}
    \put(50,-3){\footnotesize{(b)}}
    \put(15,91){\textsf{MicroBooNE $\mathsf{5.3\times10^{19}}$ POT}}
   \end{overpic}
   \begin{overpic}[width=0.329\textwidth]{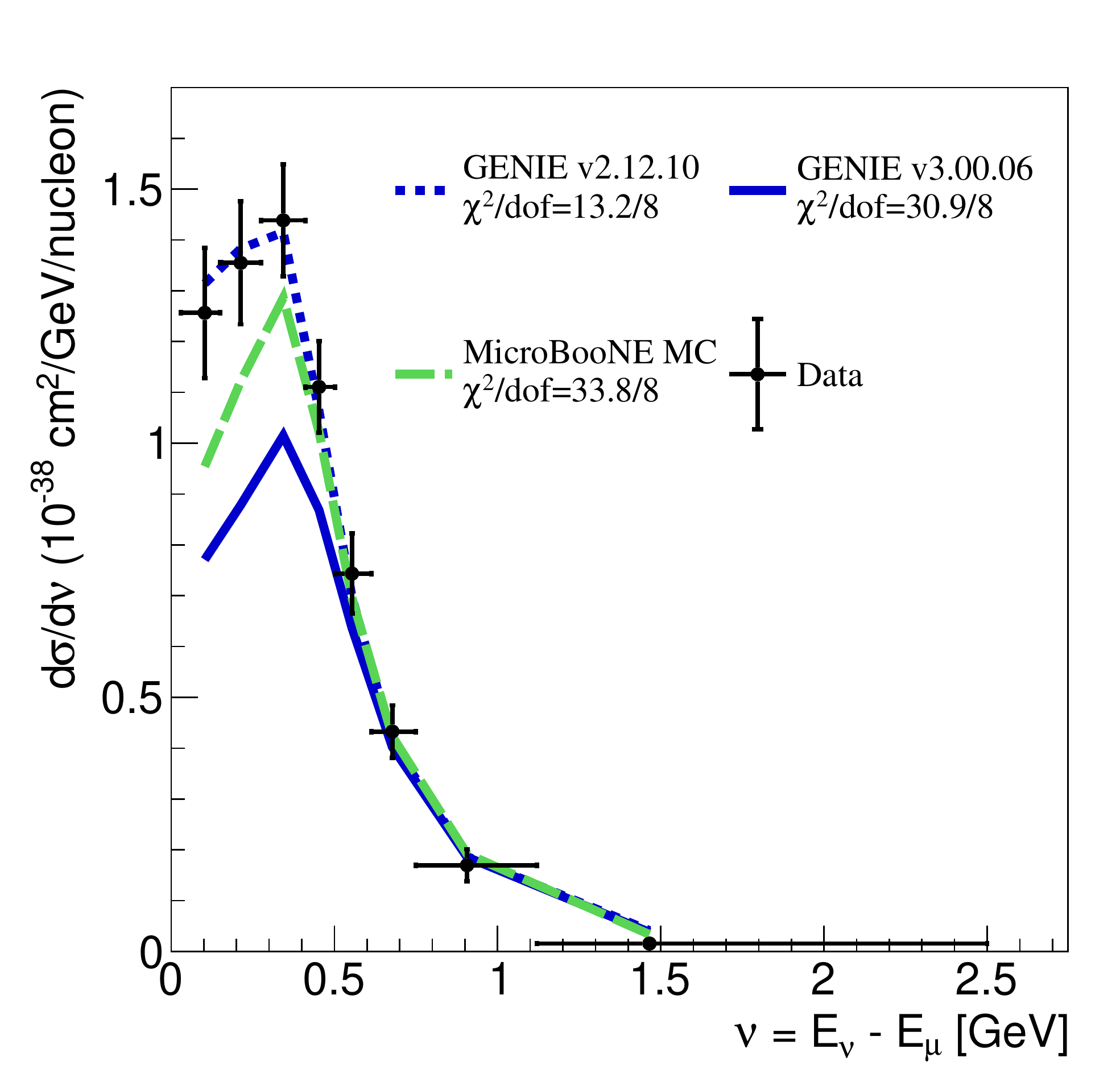}
    \put(50,-3){\footnotesize{(c)}}
    \put(16,92){\textsf{MicroBooNE $\mathsf{5.3\times10^{19}}$ POT}}
   \end{overpic}
   \caption{Comparison of the GENIE v2.12.2 default model (dashed blue) and
both the untuned (solid blue) and tuned (dashed green) GENIE v3.0.6
G18\_10a\_02\_11a models to MicroBooNE measurements~\cite{WCCCincl} of the
$\nu_\mu$-argon CC inclusive cross section as a function of (a) neutrino
energy, (b) muon energy, and (c) energy transfer. The data in plot~(a) are
divided by the bin-center neutrino energy. Chi-squared values are calculated
while neglecting uncertainties on the model predictions.}
    \label{fig:WCCCincl}
\end{figure*}

Comparisons of the GENIE v2.12.10 default model, untuned GENIE v3.0.6
G18\_10a\_02\_11a, and the ``MicroBooNE Tune'' to additional MicroBooNE
$\nu_\mu$ CC inclusive cross-section measurements from Ref.~\cite{WCCCincl} are
shown in Fig.~\ref{fig:WCCCincl}.  These data were analyzed differently than the CC inclusive in Fig.~\ref{fig:results:microMCC8} and plotted in different variables.  When the theoretical uncertainties on each
model prediction are neglected, the overall agreement between the data and the
``MicroBooNE Tune'' is found in plots (a) and (b) to be comparable to or better
than the two alternative GENIE models studied. For plot (c), the measurement of the
differential cross section as a function of energy transfer ($d\sigma/d\nu$),
the $\chi_{full}^2/\rm{N_{bins}}$ value for the tune is slightly higher than untuned
GENIE v3 and substantially higher than GENIE v2. Nevertheless, detailed studies
of the tuned model, reported in Refs.~\cite{WCCCincl,WCeLEE}, demonstrate that
it provides a reliable description of MicroBooNE data in the low-$\nu$ region
when theoretical uncertainties are taken into account.

\begin{figure}[htb]
    \centering
    \subfigure[Simulated neutrino interactions predicted by GENIE v3.0.6 G18\_10a\_02\_11a ($\chi^2_{CNP}/\rm{N_{bins}} = 32.00$/19)]{%
    \label{fig:results:ubooneDL:notune}%
	\includegraphics[width=0.5\textwidth]{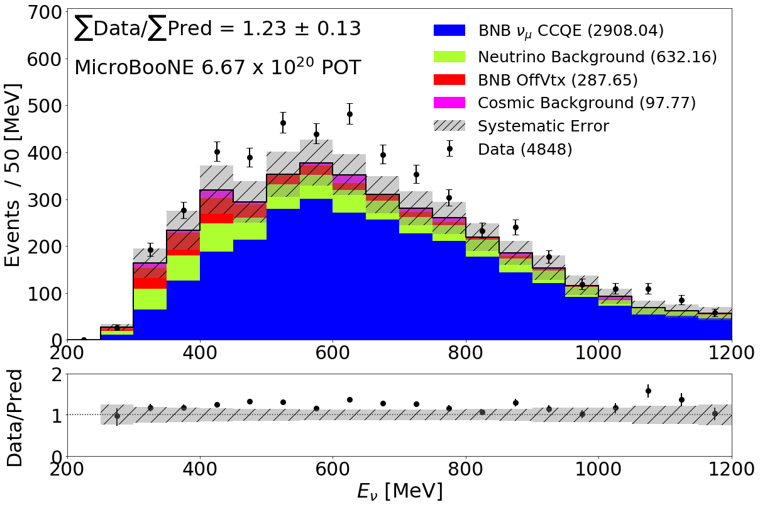}%
	}%
	\hfill
    \subfigure[Simulated neutrino interactions predicted by the ``MicroBooNE
Tune'' applied to GENIE v3.0.6 G18\_10a\_02\_11a ($\chi^2_{CNP/\mathrm{N_{bins}} = 24.96/19}$)]{
    \label{fig:results:ubooneDL:tuned}
	\includegraphics[width=0.5\textwidth]{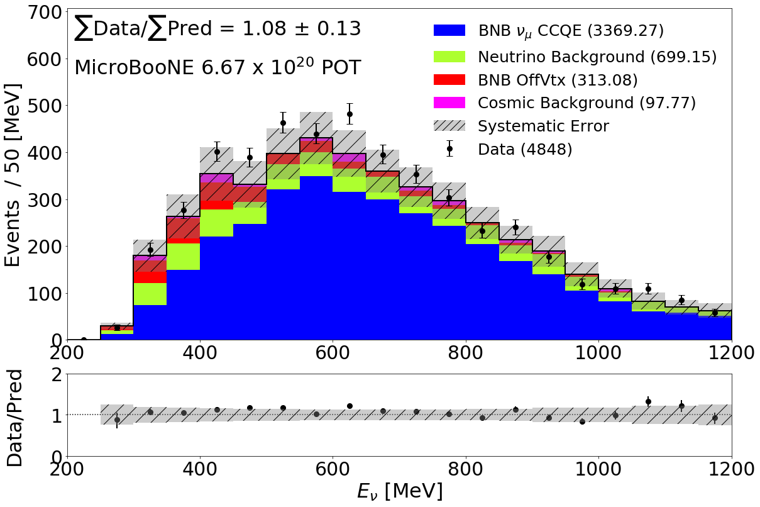}%
	}%
    \caption{Reconstructed neutrino energy of $1\mu1p$ events selected as input to a constraint for the MicroBooNE $1e1p$ LEE analysis described in Ref.~\cite{DLeLEE}. The bottom panel is taken directly from that publication and uses the ``MicroBooNE Tune'' of GENIE, while the top panel provides a similar comparison to the untuned GENIE v3.0.6 G18\_10a\_02\_11a prediction. The gray hashed region in both indicates uncertainties on the model prediction (including uncertainties on the cross-section modeling, detector modeling, and neutrino flux), and the quoted $\chi^2$ includes all uncertainties. Significantly better agreement with the data is achieved with the tuned GENIE model.}%
    \label{fig:results:ubooneDL}%
\end{figure}

As an example of the performance of the tuned model when confronted with a
MicroBooNE measurement of a highly exclusive event topology,
Fig.~\ref{fig:results:ubooneDL} compares the untuned (top panel) and tuned
(bottom panel) GENIE model predictions to the reconstructed neutrino energy
distribution obtained for the $1\mu1p$ constraint sample used in the quasielastic $\nu_e$ LEE
search~\cite{DLeLEE}. The selection used here obtains a highly pure sample of
$\nu_\mu$ CCQE events by checking the kinematic consistency of the $1\mu1p$
final state with a two-body CCQE hypothesis. Application of the ``MicroBooNE Tune''
substantially improves the level of GENIE model agreement with these data:
a ratio of $1.23 \pm 0.13$ between the integrated data and the untuned model is reduced to $1.08 \pm 0.13$ with the tune, and the combined Neyman-Pearson
chi-square ($\chi^2_\mathrm{CNP}$) metric~\cite{chi2CNP} for goodness-of-fit improves from
$\chi^2_\mathrm{CNP} = 32.00/19$ bins to $\chi^2_\mathrm{CNP}/\rm{N_{bins}} = 24.96/19$~\cite{DLeLEE}. 
Similar levels of improvement are seen for other kinematic
distributions obtained using the same data set. 
The agreement in this sample provides a valuable indication of the success of the ``MicroBooNE Tune'' in describing CCQE interactions. 

\FloatBarrier

\subsection{``MicroBooNE Tune'' Comparison to MiniBooNE Data}
\label{sec:fitresults:minib}

\begin{figure*}[htb]
    \centering
     \includegraphics[width=\textwidth]{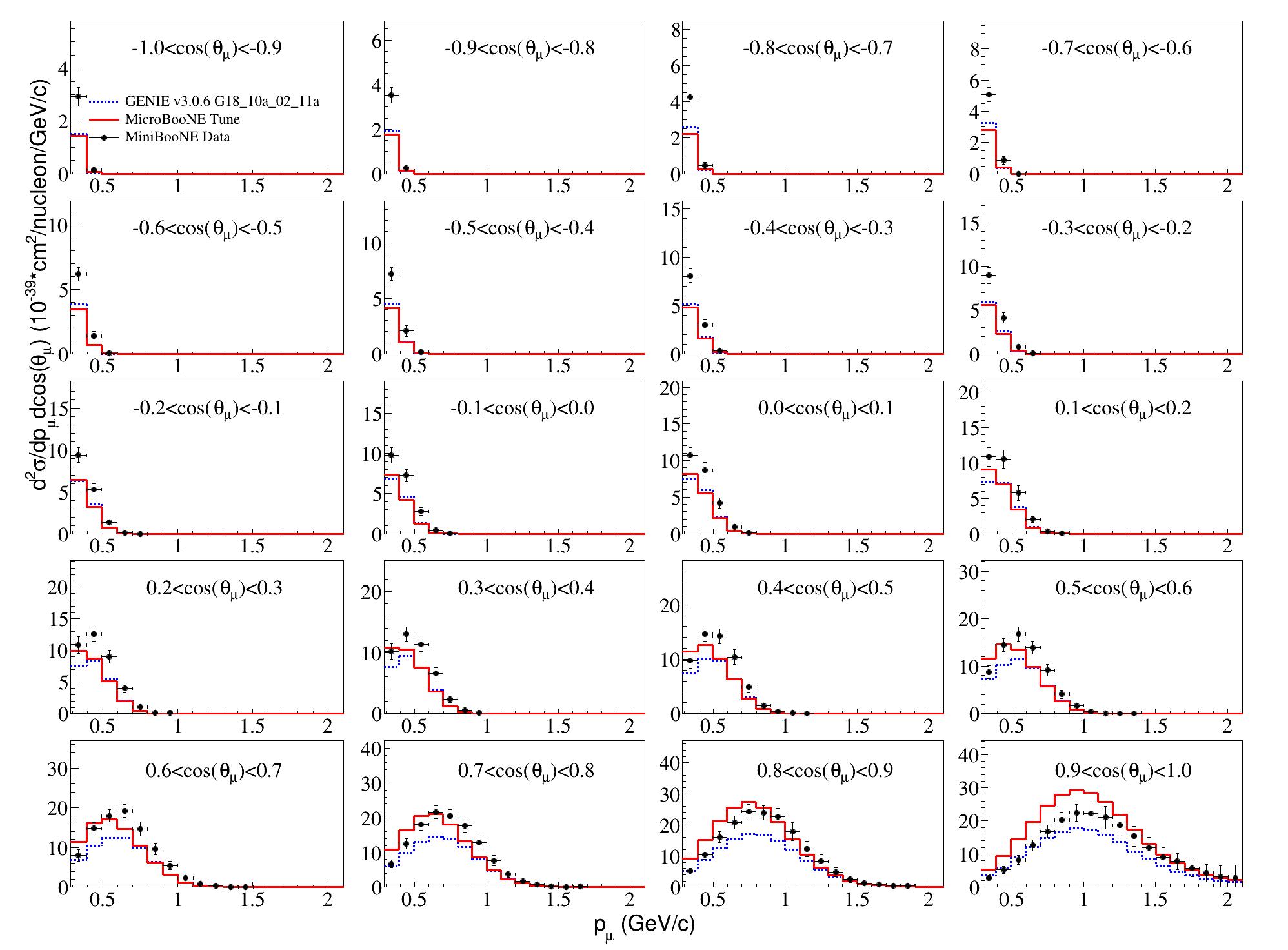}
    \caption{MiniBooNE flux-averaged CCQE-like
    double-differential cross section as a function of muon momentum and
$\rm{cos(\theta_{\mu})}$~\cite{miniboone-ccqe} compared to 
GENIE~v3 and the``MicroBooNE
Tune''. The original data release is in terms of muon kinetic energy. Uncertainties on the data points are the shape uncertainties reported by the collaboration. Table~\ref{tab:mChi2} reports the $\chi^2$ values for both the shape and normalization between the predictions and the data reported in~\cite{miniboone-ccqe}.} 
    \label{fig:results:mini}
\end{figure*}

We next compare the result of this tuning to data from the MiniBooNE
experiment.  Because this article presents a tune of the CCQE and CC2p2h
parameters, we compare to measured CCQE-like cross section
data~\cite{miniboone-ccqe}, shown in Fig.~\ref{fig:results:mini}. The
``CCQE-like'' signal definition used by the MiniBooNE collaboration
in~\cite{miniboone-ccqe} is consistent with what has been termed ``CC0$\pi$''
in this article: any interaction in which one muon, any number of nucleons, and
no other particles are produced.

Figure~\ref{fig:results:mini} shows the comparison of tuned and untuned GENIE
to the extracted CC0$\pi$ double-differential cross-section measurement from
MiniBooNE~\cite{miniboone-ccqe} as a function of lepton kinetic energy and
$\rm{cos(\theta_\mu)}$. A $\chi^2$ analysis is given in Table~\ref{tab:mChi2}.
We see overall improved agreement when using the ``MicroBooNE Tune''.
More specifically, we see very good agreement for high energy muons, but the
tuned model still underpredicts the data for low-energy muons. The original
Valencia publications~\cite{Nieves:2011yp} had very good visual agreement with
this data, although required a normalization shift of $\sim$10\%.  Using the GENIE version of the Valencia model~\cite{Gran:2013kda} and the ``MicroBooNE Tune'', no normalization shift is needed and, thus, 
the overall agreement is better.  However, the ability to describe these low
energy features is not as good as the original Valencia
model~\cite{Nieves:2011yp} despite the use of an improved binding energy
technique in GENIE.  The application of a constant binding energy in GENIE
(derived from (e,e') data at higher energies~\cite{Moniz:1971mt}) may still be
a problem. A more realistic treatment would have binding energy depending on
kinematics or the use of a spectral function, neither of which is included yet
in GENIE.

It is interesting to note that we do not see a similar underprediction when comparing to the T2K data. In fact, attempts to tune to this MiniBooNE data have
resulted in a prediction that is in tension with the T2K data, so we believe
this represents an underlying tension between the two data sets.

\begin{table*}[htb]
    \caption{$\chi^2$ for comparisons of GENIE v3.0.6 G18\_10a\_02\_11a and the ``MicroBooNE Tune'' to MiniBooNE flux-averaged CC0$\pi$-like differential cross
section data~\cite{miniboone-ccqe}, as shown in Fig.~\ref{fig:results:mini}. Following the information presented in the publication, $\chi^2$ are calculated separately for shape and normalization components, including a reported 10.7\% normalization error. Each value is then presented in the table.  No correlations between bins were included.}
    \centering
    \begin{tabular}{|c|c|c|} \hline 
    Prediction & $\chi^2_{shape}/\rm{N_{bins}}$  & $\chi^2_{norm}/\rm{N_{bins}}$ \\ \hline 
    GENIE v3.0.6 G18\_10a\_02\_11a    & 700.14/137 & 7.60/1 \\ \hline 
    ``MicroBooNE Tune''     & 323.87/137 &  1.39/1 \\ \hline  
    \end{tabular}
    \label{tab:mChi2}
\end{table*}

\FloatBarrier

\section{Interaction Model Uncertainties}
\label{sec:systematics}

The analysis presented in the preceding sections obtained refined values of
four cross-section model parameters describing the CC0$\pi$ channel, which is
dominant in MicroBooNE. A complete treatment of neutrino interaction model
uncertainties, however, requires a consideration of many additional degrees of
freedom. This section documents the full set of neutrino cross-section
systematic uncertainties adopted in current MicroBooNE analyses. With the
exception of CCQE uncertainties related to second-class currents (see
Sec.~\ref{sec:SCC}), these are implemented by adding MicroBooNE-specific
enhancements to the Reweight package~\cite{GENIErwRepo} distributed with GENIE
v3.0.6. These will be contributed by MicroBooNE for inclusion in the
upcoming v3.2.0 release of GENIE.

The GENIE Reweight code provides a standard framework for evaluating model
uncertainties via event \textit{reweighting}: the impact of a model parameter
variation on an existing sample of simulated events is approximated by
assigning a numerical weight $w$ to each event equal to the likelihood ratio
\begin{equation}
w = L^\prime / L \,.
\end{equation}
Here $L$ is the likelihood of generating the event under the original GENIE
configuration and $L^\prime$ is the corresponding likelihood calculated after
one or more interaction model parameters have been altered from their original
values. Uncertainties on distributions calculated using the events may then be
evaluated by comparing results obtained with different sets of event weights.

A typical uncertainty quantification method used by MicroBooNE is the
\textit{multiple universe} approach: a covariance matrix $V_{ij}$ describing
the uncertainty on a vector of predicted event counts $n_i$ is constructed via
the formula
\begin{equation}
V_{ij} = \frac{1}{N} \sum_{k = 1}^{N} \, (n_i^k - n_i^{CV})
  (n_j^k - n_j^{CV}) \,.
\end{equation}
Here $V_{ij}$ is the covariance matrix element of interest, $n_i^{CV}$ is the
predicted number of events in the $i$th bin under the nominal (or
\textit{central-value}) GENIE model configuration, and $n_i^{k}$ is the
corresponding event count calculated under an alternate configuration labeled
as the $k$th alternate \textit{universe}. The result is averaged over a total
of $N$ alternate universes.

MicroBooNE analyses consider a wide variety of cross-section model variations
when preparing alternate-universe predictions $n_i^k$ for use with this method.
Section~\ref{sec:par:notfit} describes new parameters (and their associated
uncertainties) developed by MicroBooNE to address missing degrees of freedom in
the reweighting tools released with GENIE v3.0.6. These are supplemented with
an external calculation of CCQE model uncertainties related to second-class
currents as described in Sec.~\ref{sec:SCC}. A full inventory of
systematic uncertainties on the GENIE-based cross-section model is then documented in Sec.~\ref{sec:systematics_budget} and Appendix~\ref{sec:partables}.

\subsection{Additional Parameters Developed for MicroBooNE Interaction
Uncertainty Evaluation}
\label{sec:par:notfit}

While current MicroBooNE analyses adopt the recommended GENIE v3.0.6 parameter
uncertainties in most cases, several new parameters were developed to assess
additional uncertainties beyond the default set. Three new parameters
(\texttt{RPA\_CCQE}, \texttt{NormCCMEC}, and \texttt{XSecShape\_CCMEC}) were
already discussed in the context of the tune to T2K CC0$\pi$ data presented in
the previous sections. The following paragraphs describe the remaining new
parameters which are used for systematic uncertainty evaluation but were
excluded from the fit to external data.

\textbf{Coulomb\_CCQE}: The Valencia CCQE model accounts for the Coulomb
interaction between the residual nucleus and the outgoing lepton using a
procedure similar to the \textit{modified effective momentum approximation}
proposed by Engel~\cite{EngelCoulomb}. This involves the use of an effective
value $k_\text{eff}$ of the outgoing lepton momentum computed via
\begin{align}
\label{eq:MEMA}
E_\text{eff} &= E - V_C(r) & k_\text{eff} = \sqrt{ E_\text{eff}^2 - m^2 }
\end{align}
where $E$ is the lepton total energy, $m$ is its mass, and $V_C(r)$ is the
local nuclear Coulomb potential at the initial radial position $r$ of the
struck nucleon. An approximate uncertainty on this correction may be assessed
by scaling the value of the nuclear Coulomb potential $V_C(r)$. The
\texttt{Coulomb\_CCQE} parameter applies this scaling with a recommended
fractional uncertainty of $\pm30\%$. The impact of the Coulomb corrections at
neutrino energies relevant for MicroBooNE is typically small.

\textbf{DecayAngMEC}: At present, all 2p2h models implemented in GENIE are
inclusive, i.e., they predict the kinematics of the outgoing lepton
only~\cite{Gran:2013kda}. A two-nucleon hadronic final state is obtained in the
simulation by recourse to a rough approximation called the \textit{nucleon
cluster model}~\cite{TeppeiMEC}. Under this approach, the leptonic
four-momentum transfer is imparted to a pair of two nucleons sampled
independently from the nuclear ground state distribution and treated as a
single object. A decay of the final-state pair is then simulated to produce two
outgoing nucleons. In the default GENIE treatment of this decay, the angular
distribution is isotropic in the rest frame of the pair. The
\texttt{DecayAngMEC} parameter adjusts this behavior. A parameter value of 0
corresponds to the default isotropic distribution, while a value of 1 yields an
angular dependence proportional to $\cos^2\theta$, where the polar angle
$\theta$ is measured with respect to the 3-momentum transfer $\mathbf{q}$. Due
to the lack of theoretical guidance for the alternate angular distribution,
this specific form was chosen as a significant deviation from isotropy which is
invariant under exchange of the two nucleons. Intermediate parameter values on
the interval $[0,1]$ linearly interpolate between the two distributions. An
uncertainty corresponding to the full difference between \texttt{DecayAngMEC}
values of 0 and 1 is adopted.

\textbf{FracPN\_CCMEC}: Charged-current 2p2h interactions may occur with an
initial pair of nucleons that share the same third component of isospin ($nn$
for neutrinos, $pp$ for antineutrinos) or that differ ($pn$). The
\texttt{FracPN\_CCMEC} parameter adjusts the fraction of these events involving
an initial $pn$ pair relative to the default prediction of the Valencia CC2p2h
model. A fractional uncertainty of $\pm20\%$ is adopted subject to the
constraint that the $pn$ fraction must lie on the interval $[0, 1]$.

\textbf{FracDelta\_CCMEC}: In contrast to other theoretical treatments of CC2p2h interactions, the Valencia calculation predicts two distinct peaks in the
joint distribution of energy and momentum transfer (see the top panel of
Fig.~\ref{fig:mecshape:truth}). These arise from two classes of Feynman
diagrams, one of which involves an internal $\Delta$ line while the other does
not. The \texttt{FracDelta\_CCMEC} parameter adjusts the relative strength of
these two contributions to the CC2p2h cross section. An uncertainty of
$\pm30\%$ on the nominal fraction of CC2p2h interactions that proceed via an
internal $\Delta$ is assessed, subject to the constraint that the fraction
remains within the interval $[0, 1]$.

Although changes to this parameter were expected to leave the total CC2p2h
event rate unchanged, numerical limitations were found that lead to an underprediction of CC2p2h events when the model had fewer relative events with an internal $\Delta$. These limitations involve
occasional very large event weights and inconsistencies in the GENIE
implementation of the cross sections themselves, e.g., the internal $\Delta$
contribution sometimes exceeds the total cross section. The main impact of
these issues on MicroBooNE analyses is expected to be a small overestimation of
the systematic uncertainty on the normalization of CC2p2h events.

\textbf{ThetaDelta2NRad}: An enhanced rate of neutrino-induced production of
single photons is a proposed explanation for the anomalous excess of
electron-like low-energy events seen by MiniBooNE \cite{MiniBooNELEE}.
Radiative decays of $\Delta$ baryons ($\Delta \to N + \gamma$) are a major
source of single photons at MicroBooNE energies and are thus of particular
interest. In addition to the existing GENIE uncertainty on the branching ratio
for radiative $\Delta$ decay, the new \texttt{ThetaDelta2NRad} parameter is
used to assess an uncertainty on the angular distribution for this process. The
implementation is similar to the one for the parameter (\texttt{DecayAngMEC})
describing the two-nucleon angular distribution in 2p2h events: a
\texttt{ThetaDelta2NRad} value of 0 corresponds to isotropic decays, while a
value of 1 yields an angular distribution proportional to $\cos^2\theta$. An
uncertainty calculated by taking the full difference between these two extremes
is adopted.

\textbf{NormCCCOH} and \textbf{NormNCCOH}: The event reweighting tools
distributed with GENIE v3.0.6 allow for variations of two parameters in the
coherent pion production (COH) cross section: the axial mass (\texttt{MaCOHpi})
and the effective nuclear radius (\texttt{R0COHpi}). However, technical
limitations in the implementation of the Berger-Sehgal COH model used in
G18\_10a\_02\_11a lead to incompatibility with those weight calculators. New
parameters which apply a constant scale factor to the CC (\texttt{NormCCCOH})
and NC (\texttt{NormNCCOH}) coherent pion production cross section are adopted
instead, each with a $\pm100\%$ uncertainty.

\textbf{NormNCMEC}: A similar normalization-only uncertainty of $\pm100\%$ is
assessed on the GENIE Empirical model of the NC 2p2h cross section.

\subsection{Second-class current form factors}
\label{sec:SCC}

Violations of charge conjugation or time-reversal symmetry in quasielastic
neutrino-nucleon scattering can give rise to new terms in the differential
cross section proportional to the \textit{second-class current} (SCC) form
factors $F_V^3$ and $F_A^3$. While the size of the SCC component of the CCQE
cross section is limited by measurements of beta decay and related processes, a
nonvanishing contribution can lead to noticeable differences in the predicted
cross sections for $\nu_e$ and $\nu_\mu$ \cite{SCCpublication}. Separate
systematic uncertainties on the possible contributions of the $F_V^3$ and
$F_A^3$ terms in the CCQE cross section are assessed for MicroBooNE analyses
via a dedicated weight calculator implemented within the LArSoft software
framework \cite{Snider2017}. Event weights are evaluated based on the ratio of
a CCQE differential cross section calculated using the NEUT~\cite{NEUT}
neutrino event generator (which includes the SCC form factors) and a similar
one calculated with GENIE (which does not). Possible SCC effects for
interaction modes other than CCQE are neglected.

%
%
%
%
%
%
%
%
%


\subsection{Final-state interaction reweighting}
\label{sec:FSIrw}

The empirical model used in GENIE v3.0.6 G18\_10a\_02\_11a for hadronic
final-state interactions is called \textit{hA2018} and represents an updated
version of the historical default treatment \textit{hA} used in GENIE
v2~\cite{FSIs,GENIEepj}. An \textit{effective cascade} approach is employed:
each hadron emerging from the primary neutrino interaction vertex may rescatter
a maximum of one time inside the nucleus, and the specific reinteraction mode
(charge exchange, inelastic, etc.) is sampled using ratios of energy- and
$A$-dependent cross sections. The hA FSI models are specifically designed to
allow for straightforward uncertainty quantification via reweighting. A key
feature of the reweighting strategy is the \textit{unitarity constraint}.  Each change is balanced by changes in other channels to make the overall fraction of events affected by FSI constant.  
Thus, inclusive event distributions (i.e., those that are not sensitive to final hadron multiplicities and kinematics) are expected to be invariant under FSI
model variations~\cite{GENIEv2Manual}.

Internal validations of the ``MicroBooNE Tune'' revealed some minor 
inconsistencies in the hA2018 reweighting tools included with GENIE v3.0.6.
These lead to violations of the aforementioned unitarity constraint
and thus to potentially inaccurate model uncertainties. The specific
issues identified in the weight calculators are
\begin{enumerate*}[label=(\arabic*)]
\item the elastic scattering reinteraction mode was removed from the FSI model
in the hA $\to$ hA2018 update, but this change is not fully applied in GENIE
Reweight;
\item the model parameters and hadron starting positions used to calculate mean free paths in Reweight are not completely identical to those used during event
generation; and
\item the mass number ($A$) and proton number ($Z$) of the nucleus are not
updated to account for prior knockout when reweighting events with multiple
primary hadrons.
\end{enumerate*}
Studies performed by all current MicroBooNE low-energy excess analyses indicate
that these problems have a negligible impact on the final results.  As with all the other additions covered in this section, this minor inconsistency is fixed in GENIE v3.2.0.

\subsection{Systematics budget}
\label{sec:systematics_budget}

Table~\ref{tab:differentpars} summarizes the ways in which the neutrino
interaction model and uncertainties used by MicroBooNE analyses differ from the
default GENIE v3.0.6 G18\_10a\_02\_11a configuration. These differences fall
into two categories: CCQE and CC2p2h parameters that were tuned as described
in this article (and therefore have different central values and uncertainties
from the default recommendation), and new parameters developed by MicroBooNE for
the evaluation of additional model uncertainties.
For completeness, a full list of parameters for which MicroBooNE analyses adopt
the GENIE recommendations unaltered is given in Appendix~\ref{sec:partables}.

{ 
\renewcommand{\arraystretch}{1.8}
    \begin{table}[htb]
        \caption{Summary of parameters for which MicroBooNE analyses adopt a
different central value and/or uncertainty than recommended in the GENIE v3.0.6
G18\_10a\_02\_11a model set.}
        \label{tab:differentpars}
    \centering
    \centerline{%
    \begin{minipage}{0.52\textwidth}
        \centering
        \begin{tabular}{l@{\hspace{0.3cm}}ccc}
        \toprule
            & \multicolumn{3}{c}{``MicroBooNE Tune''} \\
\cmidrule(lr){2-4}
            Parameter & Central value & +1$\sigma$ & -1$\sigma$ \\
            \midrule
             MaCCQE\footnote{The GENIE default value for this parameter is 0.961242 $\pm$ 0.03 GeV} & 1.10 GeV & +0.1 GeV & -0.1 GeV \\
             RPA\_CCQE\footnote{Variations are not capped at 100\%} & 85\% & +40\% & -40\% \\
             NormCCMEC & 166\%  & +50\%  & -50\%   \\
             XSecShape\allowbreak\_CCMEC & Empirical\footnote{Nominal prediction of the GENIE Empirical CC 2p2h model} & N/A & Valencia\footnote{Nominal prediction of the Valencia CC 2p2h model}  \\
             Coulomb\_CCQE & Nominal & +30\%  & -30\%   \\
             DecayAngMEC & Isotropic & Alternative\footnote{\label{foot:angle}An angular distribution proportional to $\cos^2\theta$. See the description of this parameter in Sec.~\ref{sec:par:notfit}}& N/A   \\
             FracPN\_CCMEC & Valencia & +20\% & -20\%   \\
             FracDelta\_CCMEC & Valencia & +30\% & -30\%  \\
             NormNCMEC & Nominal & +100\%  & -100\% \\
             ThetaDelta2NRad & Isotropic & Alternative\textsuperscript{\ref{foot:angle}} & N/A  \\
             NormCCCOH & Nominal & +100\% & -100\% \\
             NormNCCOH & Nominal & +100\% & -100\% \\
        \bottomrule
        \end{tabular}
    \end{minipage}
    }
    \end{table}
}

\subsection{``MicroBooNE Tune'' Total Cross Sections with Uncertainties}
\label{sec:fitresults:total_xsec}

\begin{figure}
    \caption{Predictions from the ``MicroBooNE Tune'' for the CC inclusive
total cross section for electron (top) and muon (bottom) neutrino-argon
scattering.}
    \label{fig:tune_total_xsec}
    \centering
    \includegraphics[width=0.99\columnwidth]{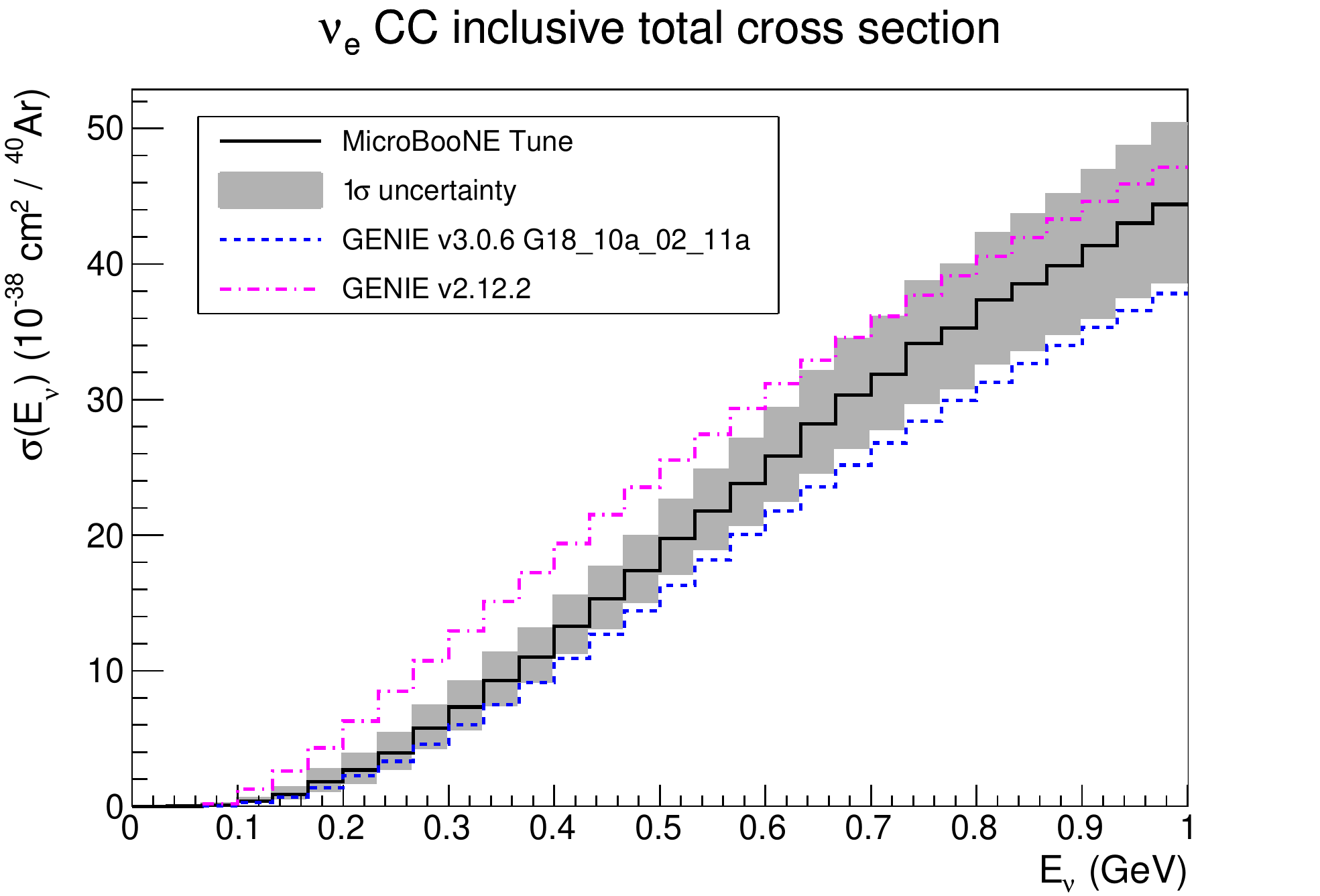}
    \vspace{\baselineskip}
    \includegraphics[width=0.99\columnwidth]{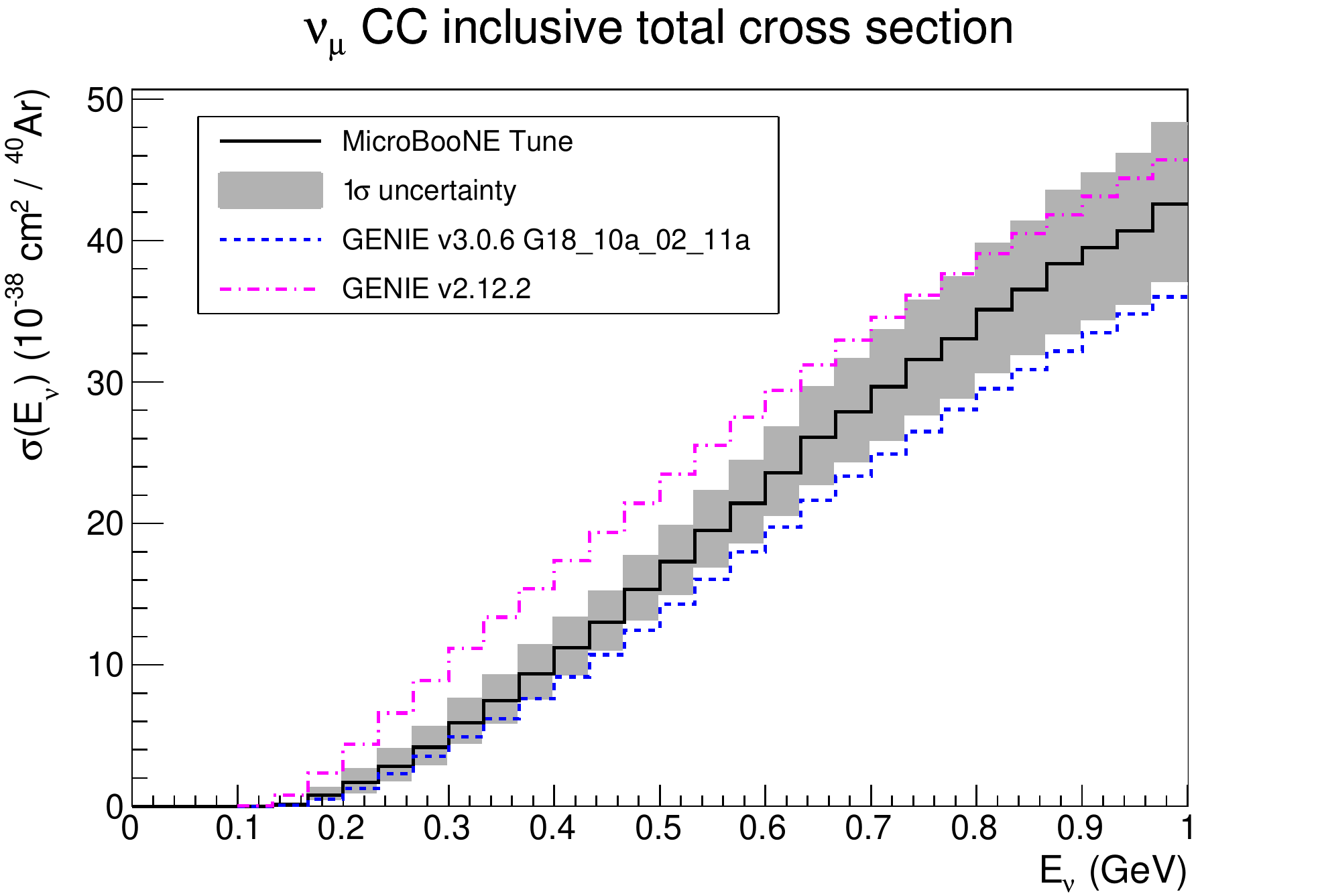}
\end{figure}

The electron-like LEE searches in MicroBooNE will search specifically for
electron neutrino interactions. Therefore it is important to also examine the
impact of the ``MicroBooNE Tune'' on the $\nu_e$ CC cross section. The modeling
of electron and muon neutrinos in GENIE uses the same underlying model
parameters, such that the tune -- while derived from fitting to muon-neutrino
cross-section data -- is also applied to the electron neutrino prediction, with
uncertainties as previously described. The top (bottom) panel of
Fig.~\ref{fig:tune_total_xsec} shows predictions for the total charged-current
cross section for electron (muon) neutrino-argon scattering obtained using the
results of our tuning procedure and full treatment of systematic uncertainties.
This is plotted as a black line as a function of neutrino energy in the region
below 1~GeV where our tune has the largest impact. The gray band indicates the
full one-sigma uncertainty for the MicroBooNE standard analysis on the
cross-section prediction. A comparison is made to the GENIE v3.0.6
G18\_10a\_\allowbreak02\_11a model used as the basis for the tune (dashed blue)
and the historical default model from GENIE v2.12.2 (dot-dashed pink).

\section{Summary and Conclusions}
\label{sec:conclusions}

This paper introduces a new tune of the GENIE event generator~\cite{GENIE} that
is suitable for MicroBooNE LEE and cross section analyses.  New parameters
designed to focus on the most uncertain parts of existing models within GENIE were developed for this purpose.  

The T2K inclusive CC0$\pi$ data~\cite{t2k2016} was the basis for this fit for a
variety of reasons. Since there was a strong desire to make this underlying
model independent of MicroBooNE data, no internal data was used in the fit.  In
addition, MiniBooNE data was excluded from the fit to avoid use of a data set
that used the same neutrino beam as MicroBooNE.  However, the MiniBooNE
CC0$\pi$ data~\cite{miniboone-ccqe} was used as a consistency check.

Initial fits with the full correlated T2K uncertainties resulted in the odd
result of a smaller $\chi^2$ and a fit cross section with significantly smaller magnitude than the data.  This result was attributed to Peelle's
Pertinent Puzzle~\cite{ppp,NuWroFit} and only the diagonal uncertainties were used for the fit presented in this work.  A number of tests of the robustness of the results were made.  Fitting exercises with alternate parameter sets and different starting values gave results very similar to those reported here.  
Peelle's Pertinent Puzzle has solutions in the literature~\cite{Pronyaev:2005cgb,kochChi2} and one of them~\cite{kochChi2} was used in an alternate fit.  The results are similar to the results reported here, with the alternate fit resulting in parameters within uncertainties of the original fit. 

The GENIE G18\_10a model set was chosen as the basis for the fit because it
provides the best description of data in this kinematic region among available
GENIE configurations.  It includes the local Fermi gas momentum distribution
and the Valencia CCQE~\cite{NievesCCQE} and CC2p2h~\cite{Gran:2013kda} models
as implemented in GENIE v3.0.6.  As discussed in the text, the implementation
in GENIE was somewhat different than what was used in
comparisons~\cite{Nieves:2011yp} to MiniBooNE data~\cite{miniboone-ccqe}.

Parameters for fitting were chosen according to relevance to the MicroBooNE
data and the level of theoretical understanding: the CCQE axial mass, the
strength of RPA corrections, the normalization of the CC2p2h cross section, and
the shape of the CC2p2h cross section. In particular, CC2p2h and RPA modeling
show significant variation among theoretical calculations~\cite{nustec-review}.  Thus, we call these theory-driven fit parameters.

When these parameters were fit to T2K CC0$\pi$ data, both the CCQE and CC2p2h normalizations were increased.  The CC2p2h shape
parameter had a weak preference for the one-peak (Empirical) shape which is
traditional in electron scattering modeling.  This sensitivity to CC2p2h shape
was not apparent in previous calculations~\cite{Nieves:2011yp}.  Although there
is a preference for a large RPA contribution as in the Valencia CCQE
model~\cite{NievesCCQE}, the constraint is not strong.  The final fit is shown
in Fig.~\ref{fig:results:fit} with the parameter correlations in
Fig.~\ref{fig:results:corr}.

The interpretation of these fit results should be carefully considered. The fit
parameter values only make sense in the context of the modeling parameters and
data set used.  Given the small number of parameters used, the results indicate
directions rather than an actual measurement.  The main fit using only diagonal elements of the covariance matrix was presented here for practical reasons.  Although it agrees well with a fit using the complete uncertainties, it is not ideal.

The ``MicroBooNE Tune'' leads to a factor-of-two reduction in the T2K data  vs. simulation chi-squared score when bin-to-bin correlations are neglected ($\chi^2_{diag}$). However, scores calculated using the more complete definitions of $\chi^2$ listed in Table~\ref{tab:fit_resultsNormShape} show that room remains for future model fitting improvements. In the present scope, performing a fit which describes T2K data optimally is less crucial than obtaining a tuned GENIE model and associated uncertainties which can be applied successfully in MicroBooNE analyses.  
As seen in Figs.~\ref{fig:results:ubooneWC}-\ref{fig:results:ubooneDL}, the results of this effort provide a noticeably improved match to data most relevant to the LEE search result~\cite{DLeLEE,WCeLEE,PeLEE,PRLeLEE}.  On the other hand, the ability to describe the MicroBooNE CC inclusive cross-section data~\cite{MCC8_CCinc} in Fig.~\ref{fig:results:microMCC8} and Table~\ref{tab:microMCC8} is degraded somewhat.  The $\chi^2$ is very large in all three cases (untuned GENIE, the ``MicroBooNE Tune'', and the alternate fit), perhaps due to the inclusion of additional processes such as pion production which were not included in the fit.  In addition, the MicroBooNE CC inclusive cross-section measurement was done with an earlier analysis package that relied on GENIE v2 and does not include recent improvements to reconstruction and analysis techniques that were used in the LEE search results~\cite{ DLeLEE,WCeLEE,PeLEE,PRLeLEE}.  More recent CC inclusive cross-section data shown in Fig.~\ref{fig:WCCCincl}, which include these improvements, typically show better agreement with the ``MicroBooNE Tune'' than with GENIE v3.0.6.   

The final fit parameters also produce a
better fit to the MiniBooNE data than GENIE v3.0.6.   However, it is not a good match to
those results. The biggest discrepancies come at the lowest muon momenta.

Nevertheless, the primary goal of producing a parameterization of relevant muon
neutrino cross-section data has been achieved. This results in an overall
improvement for the description of MicroBooNE data as discussed in
Sec.~\ref{sec:fitresults:uboone}. Since the same models are used for muon and
electron neutrino CCQE and CC2p2h interactions, this information applies to
both flavors and will provide a solid basis for MicroBooNE's searches for
physics beyond the Standard Model using electron neutrinos.  Since a fit to carbon gives a good description of argon data, the accuracy of the dependence on the nucleus ($A$-dependence) within GENIE for the relevant processes has been demonstrated.  This work will also
provide the basis for further studies of MicroBooNE cross
sections~\cite{MCC8_CCinc,uB_CCQE,uB_CCnp} and other~\cite{argoneut-xs} data
for heavier targets in the future.  

The uncertainty band derived from this work is as important as, perhaps even
more so than the central-value prediction. The cross-section model can be
further constrained experimentally through the use of background-enriched and
sideband samples, but those procedures rely on a well-motivated quantification
of systematic uncertainties in order to work effectively. In particular, this
work motivates new uncertainties on parameters for which we previously had no
or extremely weak prior uncertainties such as the strength of RPA corrections
and the modeling of CC 2p2h interactions.

\section{Acknowledgements}
\label{sec:acknowledgements}
This document was prepared by the MicroBooNE collaboration using the resources of the Fermi National Accelerator Laboratory (Fermilab), a U.S. Department of Energy, Office of Science, HEP User Facility. Fermilab is managed by Fermi Research Alliance, LLC (FRA), acting under Contract No. DE-AC02-07CH11359.  MicroBooNE is supported by the following: the U.S. Department of Energy, Office of Science, Offices of High Energy Physics and Nuclear Physics; the U.S. National Science Foundation; the Swiss National Science Foundation; the Science and Technology Facilities Council (STFC), part of the United Kingdom Research and Innovation; the Royal Society (United Kingdom); and The European Union’s Horizon 2020 Marie Sklodowska-Curie Actions. Additional support for the laser calibration system and cosmic ray tagger was provided by the Albert Einstein Center for Fundamental Physics, Bern, Switzerland. We also acknowledge the contributions of technical and scientific staff to the design, construction, and operation of the MicroBooNE detector as well as the contributions of past collaborators to the development of MicroBooNE analyses, without whom this work would not have been possible.  We are grateful to the NUISANCE team for extra support for the software tools used in this study.

\appendix
\section{List of Parameters and Uncertainties}
\label{sec:partables}

Table~\ref{tab:dials_default} lists all model parameters from GENIE v3.0.6
G18\_10a\_02\_11a for which the central value and uncertainty are adopted
unaltered as part of the ``MicroBooNE Tune'' described in this paper.
Parameters which have been added or altered by MicroBooNE are listed
separately in Table~\ref{tab:differentpars}.

\makeatletter
\newenvironment{xlongtable}[1]
 {\appdef\table@hook{#1}\longtable}
 {\endlongtable}
\makeatother

{ 
\renewcommand{\arraystretch}{1.8}
    \begin{xlongtable}{\footnotesize}{lccc}
               \toprule
            Parameter & Central value & +1$\sigma$ & -1$\sigma$ \\
            \midrule
             \label{tab:dials_default}  
            \endfirsthead
            \endhead
            \bottomrule
            \endlastfoot
            \endfoot
            \multicolumn{4}{l}{ \textbf{CCQE form factor parameterization} } \\
             AxFFCCQEshape & Dipole & Z-expansion & N/A \\
             VecFFCCQEshape & BBA07 & Dipole & N/A \\
            \multicolumn{4}{l}{ \textbf{NC elastic form factors} } \\
            MaNCEL & 0.961242 GeV & +25\%  & -25\%  \\
            EtaNCEL & 0.12 & +30\%  & -30\%  \\
            \multicolumn{4}{l}{ \textbf{RES form factors and decays} } \\
            MaCCRES & 1.065047 GeV & +20\%  & -20\%  \\
            MvCCRES & 0.840 GeV & +10\%  & -10\%  \\
            MaNCRES & 1.120 GeV & +20\%  & -20\%  \\
            MvNCRES & 0.840 GeV & +10\%  & -10\%  \\
            RDecBR1gamma & Nominal & +50\% & -50\% \\
            RDecBR1eta & Nominal & +50\% & -50\% \\
            Theta\_Delta2Npi & Nominal & Isotropic & N/A \\
            \multicolumn{4}{l}{ \textbf{AGKY hadronization model} } \\
            AGKYxF1pi & -0.385 & +20\%  & -20\%  \\
            AGKYpT1pi & $1 / 6.625$ GeV$^2$ & +3\%  & -3\%  \\
            \multicolumn{4}{l}{ \textbf{Normalization of non-RES final states} } \\
            NonRESBGvpCC1pi & 0.007713 & +50\%  & -50\%  \\
            NonRESBGvpCC2pi & 0.787999 & +50\%  & -50\%  \\
            NonRESBGvnCC1pi & 0.127858 & +50\%  & -50\%  \\
            NonRESBGvnCC2pi & 2.11523 & +50\%  & -50\%  \\
            NonRESBGvbarpCC1pi & 0.127858 & +50\%  & -50\%  \\
            NonRESBGvbarpCC2pi & 2.11523 & +50\%  & -50\%  \\
            NonRESBGvbarnCC1pi & 0.007713 & +50\%  & -50\%  \\
            NonRESBGvbarnCC2pi & 0.787999 & +50\%  & -50\%  \\
            NonRESBGvpNC1pi & 0.1 & +50\%  & -50\%  \\
            NonRESBGvpNC2pi & 1 & +50\%  & -50\%  \\
            NonRESBGvnNC1pi & 0.3 & +50\%  & -50\%  \\
            NonRESBGvnNC2pi & 1 & +50\%  & -50\%  \\
            NonRESBGvbarpNC1pi & 0.3 & +50\%  & -50\%  \\
            NonRESBGvbarpNC2pi & 1 & +50\%  & -50\%  \\
            NonRESBGvbarnNC1pi & 0.1 & +50\%  & -50\%  \\
            NonRESBGvbarnNC2pi & 1 & +50\%  & -50\%  \\
            \multicolumn{4}{l}{ \textbf{Bodek-Yang structure functions} } \\
            AhtBY & 0.538 GeV$^2$ & +25\%  & -25\%  \\
            BhtBY & 0.305 GeV$^2$ & +25\%  & -25\%  \\
            CV1uBY & 0.291 GeV$^2$ & +30\%  & -30\%  \\
            CV2uBY & 0.189 GeV$^2$ & +40\%  & -40\%  \\
            \multicolumn{4}{l}{ \textbf{Final-state interactions} } \\
            MFP\_pi & hA2018 & +20\%  & -20\%  \\
            MFP\_N & hA2018 & +20\%  & -20\%  \\
            FrCEx\_pi & hA2018 & +50\%  & -50\%  \\
            FrInel\_pi & hA2018 & +40\%  & -40\%  \\
            FrAbs\_pi & hA2018 & +30\%  & -30\%  \\
            FrPiProd\_pi & hA2018 & +20\%  & -20\%  \\
            FrCEx\_N & hA2018 & +50\%  & -50\%  \\
            FrInel\_N & hA2018 & +40\%  & -40\%  \\
            FrAbs\_N & hA2018 & +20\%  & -20\%  \\
            FrPiProd\_N & hA2018 & +20\%  & -20\%  \\
    \caption{\small GENIE model parameters
used with default settings in the ``\mbox{MicroBooNE} Tune.''}
    \end{xlongtable}
}

%

\end{document}